\documentclass[fleqn,usenatbib]{mnras}

\usepackage{newtxtext,newtxmath}

\usepackage[T1]{fontenc}

\DeclareRobustCommand{\VAN}[3]{#2}
\let\VANthebibliography\thebibliography
\def\thebibliography{\DeclareRobustCommand{\VAN}[3]{##3}\VANthebibliography}

\usepackage{graphicx}    
\usepackage{amsmath}

\title[Implications of the MW's contracted DM halo]{Implications of a contracted dark matter halo for the Milky Way's inferred virial mass}

\author[D. Dado et al.]{
Diego Dado,$^{1,2}$\thanks{E-mail: diego.dado2@durham.ac.uk}
Shaun T. Brown,$^{1,3}$
Azadeh Fattahi,$^{1,3}$
Andreea S. Font$^{4}$
and Ian G. McCarthy$^{4}$
\\
$^{1}$ Institute for Computational Cosmology, Department of Physics, Durham University, South Road, Durham DH1 3LE, UK\\
$^{2}$ Centre for Extragalactic Astronomy, Department of Physics, Durham University, South Road, Durham DH1 3LE, UK\\
$^{3}$ The Oskar Klein Centre, Department of Physics, Stockholm University, Albanova University Center, 106 91 Stockholm, Sweden\\
$^{4}$ Astrophysics Research Institute, Liverpool John Moores University, 146 Brownlow Hill, Liverpool L3 5RF, UK
}

\date{Accepted XXX. Received YYY; in original form ZZZ}

\pubyear{2026}

\begin{document}
\label{firstpage}
\pagerange{\pageref{firstpage}--\pageref{lastpage}}
\maketitle

\begin{abstract}
We investigate how reliably the global properties of Milky Way–mass dark matter haloes can be recovered from dynamical data from a limited range of radii, particular $\lesssim 30~\mathrm{kpc}$ where observations are most sensitive but baryonic processes modify the halo structure. Using the ARTEMIS simulations, which produce varying degrees of baryon-induced contraction, we fit dark matter profiles over restricted radial ranges using commonly adopted parametric models. Assuming negligible observational uncertainties allows the systematic errors from these choices to be isolated. When fits are confined to inner radii, an NFW profile underestimates the virial mass by a factor of $\approx 2$ on average ($\approx4$ for some systems), and the concentration by a factor of $\approx2$. Einasto and generalised-NFW models provide excellent local fits but retain similar global biases. In contrast, a contracted halo prescription yields stable extrapolations and recovers unbiased halo mass estimates over all radii. The inferred mass improves systematically with increasing radial coverage, and tracers beyond $\gtrsim 50 \, \mathrm{kpc}$ largely eliminate the mean bias for all models. The local dark matter density at the Solar radius is recovered to within $\lesssim 5 $\% for all profiles other than NFW. These biases are sufficient to reconcile recent low Milky Way mass estimates derived from inner rotation-curve analyses with the canonical $\approx 10^{12} \mathrm{M}_\odot$. We additionally find a halo-to-halo scatter of $\gtrsim 0.1$ dex ($\approx 25$\%) persists even under idealised conditions, setting a likely lower limit for the precision of halo mass estimates.
\end{abstract}

\begin{keywords}
cosmology: dark matter -- galaxies: kinematics and dynamics -- galaxies: evolution
\end{keywords}

\section{Introduction}

In the $\Lambda$CDM cosmogony, galaxies form at the centres of virialised dark matter (DM) haloes, with their subsequent evolution shaped by the interplay between hierarchical accretion, gas cooling, and feedback from stars and active galactic nuclei \citep[e.g.][]{PressSchecter1974, Binney_77, Frenk_88, cimatti2019, Marasco_2021}. The mass and internal structure of a galaxy’s host halo are therefore inextricably linked to its own formation and evolution.

Observationally, we can infer the distribution of DM within galaxies through the kinematics of the stars and gas within different components of a galaxy, with such analysis providing some of the strongest early evidence for the existence of DM \citep[e.g.][]{Rubin_80}. Typically, for nearby galaxies similar to the Milky Way, the most robust kinematic data comes from the disk ($\lesssim  30\, \mathrm{kpc}$). With dedicated surveys we now have robust kinematic data for large samples of galaxies in the nearby Universe \citep[e.g.][]{Persic_96, Things, Little_things}. Even within this context, observations within the Milky Way (MW) offer unparalleled detail to map the DM distribution within our host halo \citep[e.g. reviews by][]{Bullock_2017, zavala2019}. In addition to the quality of data within the Milky Way, we also have access to a range of kinematic probes beyond the central disk, including the stellar halo \citep[e.g.][]{Xue_2008, Deason_2012, Kafle_2012}, globular clusters \citep[e.g.][]{Eadie_2017, Posti_2019, Watkins_2019} and satellite galaxies \citep[e.g.][]{Watkins_2010, Li_2017, Patel_2018, Callingham_2019}.

Reliable theoretical models for the distribution of mass with DM are necessary to be able to robustly interpret kinematic data. In the absence of baryonic effects it is well established that haloes follow on average a Navarro–Frenk–White \citep[NFW;][]{Navarro_1997} profile, which exhibits an inner density slope of $\rho \propto r^{-1}$ and an outer slope of $\rho \propto r^{-3}$, with a characteristic scale radius, $r_{\mathrm{s}}$, describing the transitions between the two limits. An NFW profile has two free parameters, typically phrased as the halo mass and concentration, defined as the ratio of the virial to scale radius. Cosmological simulations predict these to be strongly correlated leading to a well defined concentration-mass relation that varies with redshift and cosmology \citep[e.g.][]{Dutton_2014, Diemer, Ludlow_2016, Brown_20}. These models predict a Milky Way mass halo at $z=0$ to have a mean concentration of $c \approx 8$, with a scatter of $\approx 0.1 \, \mathrm{dex}$, where the concentration is defined relative to a spherical overdensity radius $R_{200\rm c}$ -- i.e., the radius enclosing a mean matter density 200 times the critical density at $z=0$.

While the internal structure of DM haloes in the absence of baryons is relatively well understood, it has been shown that the inclusion of baryonic processes can significantly impact the central structure of the DM halo. In some cases, strong feedback-driven gaseous outflows can lead to the formation of a DM core due to energy transfer from a time-varying gravitational potential \citep[e.g.][]{Pontzen_12, Di-Cintio_14}. The efficiency of this process is strongly mass dependent, with the largest cores forming at intermediate stellar-to-halo mass ratios, $M_\star/M_{\rm halo} \sim 10^{-2}$--$10^{-3}$, and diminishing both at lower and higher masses \citep[e.g.][]{Di-Cintio_14, Lazar2020}. Recent work has further shown that this behaviour is closely linked to the burstiness of star formation, helping to reconcile results across different simulation suites and to understand the halo-to-halo scatter in such formalisms \citep[][]{DREAMS2026, Sarrato-Alos2026}. For systems of Milky Way mass or larger, many simulations instead predict that the central concentration of baryons within the galaxy can lead to a contracted DM halo, with an increased inner density relative to a DM-only simulation -- an effect arising from principles of action conservation and gas cooling when baryons are condensing at the centre of such halos \citep[e.g.][]{Blumenthal1986, Gnedin_2004, Schaller_15, Poole-McKenzie_20, Callingham_20}. 

This significantly complicates the task of inferring the host DM halo properties from galactic dynamics, as the host halo not only influences the evolution of the galaxy, but the galaxy also influences the halo itself. The most precise dynamical constraints are often obtained from the disk of a galaxy, where baryons dominate the potential, and therefore from the same region where the impact of baryons on the DM halo is expected to be most significant. Consequently, without a reliable theoretical model to account for baryonic effects, inferences of DM halo properties may be subject to significant systematic errors.

In recent years, a tension has emerged in estimates of the Milky Way halo mass between measurements based on dynamics in the central Galaxy ($5 < r/\mathrm{kpc} \lesssim 30$) and those using tracers in the outer halo ($r \gtrsim 30 \mathrm{kpc}$). Studies using large, precise data sets from \textit{APOGEE} DR17 \citep{Majewski_2017} and \textit{Gaia} DR3 \citep{Gaia_2023}, but restricted to the disk, have inferred masses of $\approx 10^{11}$--$5 \times 10^{11} \, \mathrm{M_{\odot}}$, depending on the assumed profile \citep[e.g.][]{Eilers_2019, Ou_2023}, while analyses using tracers beyond the disk typically find $\approx 10^{12} \, \mathrm{M_{\odot}}$ \citep[e.g.][]{Watkins_2019, Callingham_2019} \footnote{In this context, “halo mass” or “mass” typically refers to a spherical overdensity definition such as $M_{200\mathrm{c}}$ or $M_{\mathrm{vir}}$; these definitions differ at the $\sim$10–20\% level and are therefore not sufficient to explain the discrepancy.}. Reconciling this discrepancy is crucial for a broad range of applications, from predictions of Local Group satellite abundances to the local DM density relevant for direct-detection experiments \citep{Goodman_Witten_1985, Drukier_1986, Jungman_1996}.

Addressing this problem requires a statistically meaningful sample of Milky Way-mass haloes simulated at sufficiently high resolution to robustly resolve the inner dark matter distribution, while simultaneously forming realistic central galaxies. In particular, the simulations must capture the response of the dark matter halo to baryonic assembly and feedback, which depends sensitively on both numerical resolution and the calibration of subgrid physics. To this end, we use the ARTEMIS suite of high-resolution cosmological zoom-in hydrodynamical simulations of Milky Way-mass haloes \citep{ARTEMIS}. ARTEMIS comprises a large sample of haloes with $M_{200\mathrm{c}} \sim 10^{12}\,\mathrm{M}_\odot$, simulated at baryonic mass resolution $m_{\mathrm{b}} \simeq 2 \times 10^{4}\,\mathrm{M}_\odot$, and employs a stellar feedback model calibrated to reproduce the median stellar mass-halo mass relation inferred from abundance matching applied to large extragalactic surveys. As such, ARTEMIS is not tuned to reproduce the detailed properties of the Milky Way specifically, but instead provides a representative population of Milky Way-mass galaxies suitable for assessing systematic biases in halo inference methods. By fitting the dark matter profiles of ARTEMIS haloes over a range of radii and assuming a variety of parametric halo models commonly used in the literature, including the contracted halo model of \defcitealias{Cautun_2020}{C20} Cautun et al.\ (\citeyear{Cautun_2020}, hereafter \citetalias{Cautun_2020}), we quantify how accurately halo mass and concentration can be recovered when only kinematic tracers over a limited radial range are available.

The paper is structured as follows. Section~\ref{sec:Methods} summarises the ARTEMIS simulations, the dark-matter-rescaling procedure, and our fitting methodology. Section~\ref{sec:Contraction} quantifies the impact of baryonic contraction on the inner dark matter distribution and verifies the \citetalias{Cautun_2020} power law relation in ARTEMIS. Section~\ref{sec: Response of different profiles} presents the performance and characteristic behaviour of each halo model as a function of radial fitting extent, both in local/internal metrics and in global/extrapolated metrics. Section~\ref{sec:Conclusion} summarises our proposed interpretation of these findings, especially the implications for Milky Way (and analogues) mass and concentration measurements.

\section{Methods}
\label{sec:Methods}
We use the ARTEMIS cosmological hydrodynamical simulations to study how Milky Way–mass galaxies alter their host dark matter haloes. In this section we outline the details of the simulations, the halo mass definitions used, profile construction, and fitting methodology.

\subsection{The ARTEMIS simulations}
\label{sec:ARTEMIS}
ARTEMIS (Assembly of high-ResoluTion Eagle simulations of MIlkyWay-type galaxieS; \citealt{ARTEMIS}) is a suite of 42 cosmological zoom-in hydrodynamical simulations of Milky Way-mass haloes. These simulations are run using the Gadget-3 code \citep[last described in][]{Springel_2005} with the same hydrodynamics solver and galaxy formation (subgrid) physics as the EAGLE project \citep{Schaye_2014}, but at a significantly increased resolution. This increase required a recalibration of key subgrid modules, most notably stellar feedback, to reproduce the median stellar mass-halo mass relation at $z=0$ for Milky Way-mass systems; a detailed description of the simulation setup and calibration is presented in \citet{ARTEMIS}. The recalibration was not intended to match the detailed properties of the Milky Way specifically, but rather to yield a representative population of Milky Way-mass galaxies. As such, ARTEMIS is well suited for studying generic baryonic effects on dark matter haloes -- including baryonic contraction -- and for quantifying systematic biases in halo property inference from kinematic data confined to the inner regions of galaxies.

The ARTEMIS simulations adopt a flat $\Lambda$CDM cosmology consistent with WMAP9 \citep{Hinshaw_2013}, with parameters: $\Omega_\mathrm{m} = 0.2793$, $\Omega_\mathrm{b} = 0.0463$, $h = 0.70$, $\sigma_8 = 0.8211$, and $n_\mathrm{s} = 0.972$. The simulations are run in a parent volume of $25 \mathrm{Mpc}/h$ per side, with the gravitational softening length set to $125 \mathrm{pc}/h$ for all particles, and with a baryon particle mass of $2.23 \times 10^4\mathrm{M}_\odot/h$.

In this work, we focus exclusively on the central haloes of the ARTEMIS suite, which have virial masses in the range $0.8 \leq M_{200\mathrm{c}}/10^{12}\mathrm{M}_\odot \leq 2$. This yields a sample of 42 central haloes.

\subsection{Halo finding \& virial parameters}
\label{sec:virial_definitions}

\begin{figure}
    \includegraphics[width=0.475\textwidth]{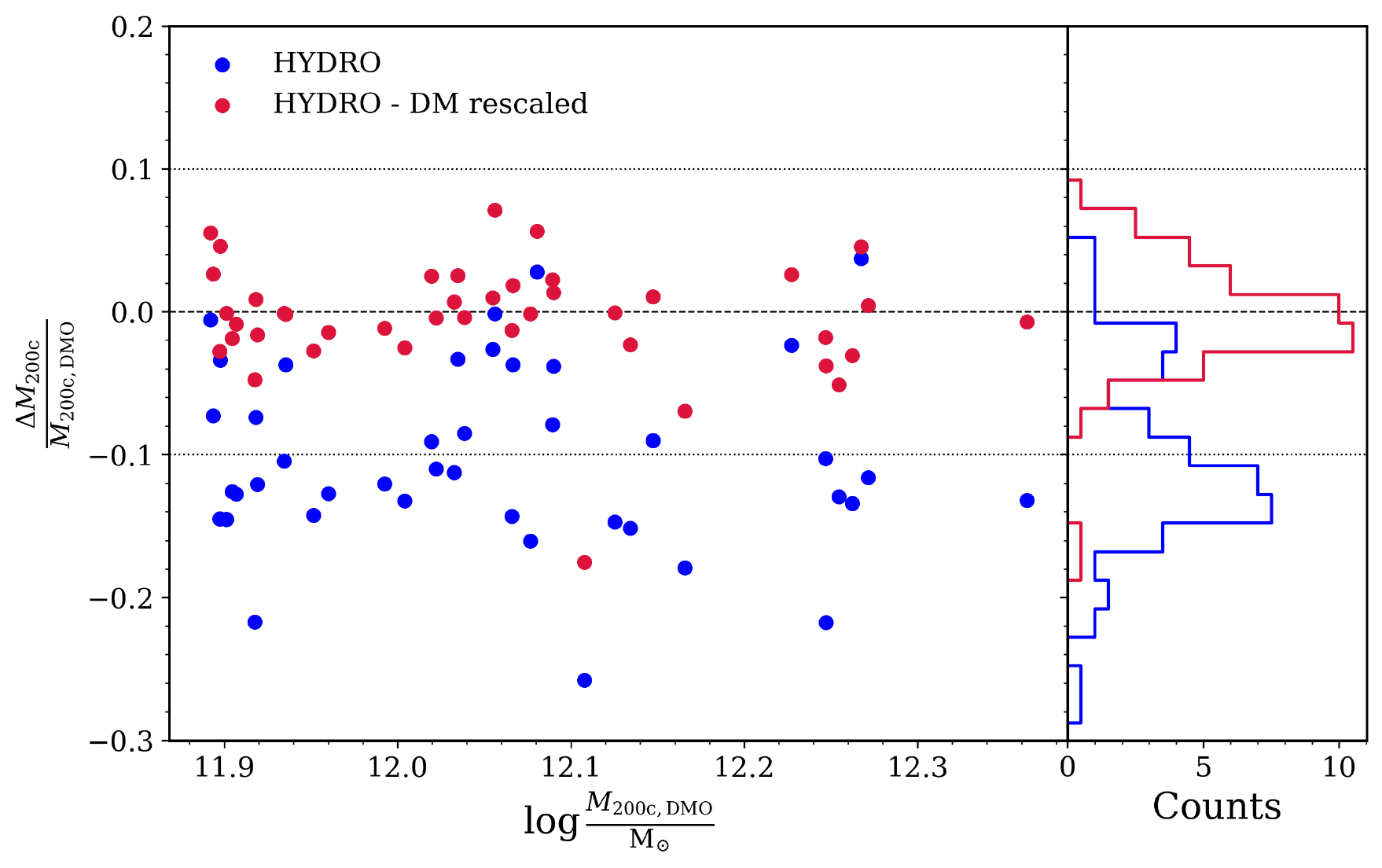}
	\caption{Fractional difference in virial mass, $(M_{200\mathrm{c}} - M_{200\mathrm{c},\mathrm{DMO}})/M_{200\mathrm{c},\mathrm{DMO}}$, between the hydrodynamical and DMO counterparts of each ARTEMIS central halo. Colours denote the definition of $M_{200\mathrm{c}}$: dark blue for the standard total enclosed mass definition (Equation \ref{eq:virial_condition}), and dark red for the dark matter–rescaled definition which scales the enclosed DM mass by the cosmic mean baryon fraction (Equation \ref{eq:mass_def_scaled}).
	\textit{Left panel:} Fractional mass differences as a function of the DMO virial mass; each circle corresponds to a single central halo.
	\textit{Right panel:} Distribution of these fractional differences.
	In both panels, the dashed black line marks perfect agreement between hydrodynamical and DMO masses, while the grey dotted lines indicate $\pm 10~\%$ deviations. Applying the DM-rescaled definition greatly suppresses the systematic offset, showing that the dominant contribution to the total mass mismatch arises from baryons displaced from the halo rather than from any substantial loss of dark matter.}
	\label{fig:Rescaling}
\end{figure}

Haloes are identified using the friends-of-friends (FOF) algorithm \citep[][]{Davis_1985} with a linking length 0.2 times the mean particle separation, and they are further split into gravitationally bound structures using the SUBFIND algorithm \citep[][]{Springel_2001}.
As is common, we use a spherical overdensity definition for the mass and size of a halo. In this approach, the halo is bounded by a radius $R_{\Gamma}$ within which the mean enclosed density equals $\Gamma$ times the critical density of the Universe, $\rho_\mathrm{c}$. Formally,
\begin{equation}
\frac{3 M(<r)}{4 \pi \,r^3} = \Gamma \rho_\mathrm{c},
\label{eq:virial_condition}
\end{equation}
where $M(<r)$ is the total mass enclosed, $\rho_\mathrm{c}(z) = 3H^2(z)/8\pi G$ is the critical density at redshift $z$, and $H(z)$ is the redshift-dependent Hubble parameter. The mass and radius that satisfy this relation are then defined as halo's mass and radius, $M_{\Gamma}$ and $R_{\Gamma}$. Throughout this work we adopt an overdensity threshold of $\Gamma = 200$ relative to the critical density. The corresponding halo mass and radius are denoted as $M_{200\mathrm{c}}$ and $R_{200\mathrm{c}}$. It is very common to refer to these definitions simply as the virial mass and radius of the halo, and we will use this terminology throughout to refer to the above definition.

\subsubsection{DM rescaling}
\label{sec:DM rescaling}

In the halo mass definition described above it is most common to use the \textit{total} enclosed mass. In Fig.~\ref{fig:Rescaling} we show the difference in halo mass using the standard definition comparing the ARTEMIS haloes in a DM-only simulation with full hydrodynamics (see blue data points). For the DM-only simulations all matter is in the form of DM while for the hydrodynamic simulation the standard halo mass definition includes both DM and baryons. When comparing these simulations this definition of the halo mass leads to the baryonic simulation having systematically less massive haloes by $\approx 15 \%$, with significant scatter between systems.

One of the limitations of using all matter to define a halo's mass is that it becomes unclear if this discrepancy is due to changes in the distribution of DM, baryons or both. In this work we are primarily focussed on the effect of the baryons on distribution of DM, and less on the distribution of baryons themselves. Therefore, to isolate the changes to the DM distribution we choose to modify the halo mass definition to the following,
\begin{equation}
	\frac{1}{1-f_{\mathrm{b}}} \frac{3 M_{\mathrm{DM}} (<r)}{4\pi\, r^3} = \Gamma \rho_{\mathrm{c}},
\label{eq:mass_def_scaled}
\end{equation}
where $f_{\mathrm{b}}$ is the cosmic baryon fraction, which for our assumed cosmology is $f_{\mathrm{b}} = 0.17$. This definition is similar in form to the standard choice (Eqn.~\ref{eq:virial_condition}), and still corresponds to a spherical overdensity mass definition, but differs by using only the distribution of DM, scaled according to the baryon fraction. In a DM-only simulation ($f_{\mathrm{b}}= 0$) the two definitions are identical.

In Figure~\ref{fig:Rescaling} we show the halo mass difference with this modified definition using only the DM (red points). This halo mass definition largely removes the systematic offset in virial mass seen when using the standard convention, with the distribution having no average systematic difference and a scatter of $\approx 5 \%$. This highlights that the majority of the virial mass difference between hydrodynamical and DMO runs arises from baryons being expelled from the halo, rather than a substantial change DM enclosed within the outer halo around near the virial radius.\footnote{This result holds for our sample of exclusively Milky Way mass haloes, but may not generalise to all mass scales. For example it has been shown that for dwarf galaxies there is notable change to the DM halo mass \citep[e.g.][]{Sawala_15}.} Consequently, the dark matter rescaled definition of virial parameters provides a robust framework for comparing halo properties between hydrodynamical and DMO runs, isolating differences which reflect purely structural changes in the dark matter distribution.

Throughout the rest of the paper we only present results using the mass definition using the rescaled DM distribution (Eqn.~\ref{eq:mass_def_scaled}). For brevity we will use the terminology `halo mass', `virial mass' and $M_{\mathrm{200c}}$ to refer to this definition. We also note that one outlier halo in Figure~\ref{fig:Rescaling}, exhibiting a systematic offset of $\sim -17\%$ under the dark matter–rescaled mass definition, is removed from the sample for the rest of the paper, and is a particular unrelaxed system with an ongoing major merger.

\subsection{Circular velocity fitting \& theoretical DM profiles}
\label{sec:fitting_pipeline}
To quantify how well different theoretical halo models reproduce the simulated dark matter structure, we fit parametric profiles to the radial mass distributions of each halo.

We calculate the spherically averaged density, enclosed mass, and circular velocity radial profiles for all haloes in our sample. Throughout, we show results using $N_{\rm bins}=128$ evenly spaced logarithmic bins in radius spanning $0.21 \leq r/\mathrm{kpc} \leq 320$. Roughly corresponding to between $10^{-3}$ and $1.5$ times the typical virial radius (the median virial radius of the sample is $R_{\mathrm{200c}} = 214 \, \mathrm{kpc}$). We have tested the impact of the number of radial bins by repeating our analysis with $N_{\rm bins}=32, 64, 256$; we find broadly similar results when fitting enclosed mass and circular velocity profiles, and only mild differences when fitting the density profile. We adopt 128 bins, as this is the regime where the fitting results converge for all profiles, both in local estimates and extrapolated quantities.

We obtain the best-fit parameters and associated uncertainties for each halo model through least-squares optimisation. Specifically, we minimise
\begin{equation}
	\sigma_{\text{fit}} =\frac{1}{N-1}\sum_{i}^N (\log M(<r_i) - \log M_{\text{model}}(<r_i))^2,
    \label{sec:sigma_fit}
\end{equation}
where the sum runs over the radial bins $r_i$ used in the fitting procedure. Here $M(<r)$ is the enclosed mass measured from the simulation -- $M_{\rm DM, DMO}(<r)$ and $M_{\rm DM,\,rescaled}(<r)$ for the DMO and hydrodynamical runs, respectively -- and $M_{\text{model}}(<r)$ is the model prediction. Although we perform our fits using the enclosed mass profile, we have verified that fitting the circular velocity profile $V_c(r)$ -- which more closely resembles typical observational probes -- yields consistent results and does not affect our conclusions.  

Furthermore, while the outer radius used in the fitting procedure will be varied, we fix the lower bound to exclude regions where numerical effects can lead to biased estimates of the circular velocity \citep[see e.g.][]{Power_2003}. Our fixed lower bound is given by $r_{\text{min.~fit}} = 3 ~ \text{max}(r_{\text{conv}})$. The convergence radius, $r_{\text{conv}}$, is calculated following the prescription of \cite{Ludlow_2019}, and depends on the local particle number and enclosed volume. Although all simulations share the same mass resolution, the convergence radius varies between haloes owing to differences in their internal structure and particle distributions. We therefore adopt a conservative threshold defined as three times the maximum $r_{\text{conv}}$ across our halo sample -- which corresponds to $\approx 3 \, \mathrm{kpc}$ for our haloes. Note that some of the most recent works targeting an estimate of the Galactic halo mass adopt a cut at $r \approx 5~\mathrm{kpc}$, due to the complexity of modelling the strongly non-axisymmetric potential induced by the Galactic bar \citep[e.g.][]{Ou_2023}; hence, our cut at $\approx 3~\mathrm{kpc}$ is broadly comparable to those adopted in observational analyses.

Having defined our fitting procedure, we now introduce the analytic dark matter density profiles used in this work.

\subsubsection{NFW}
\label{sec:NFW}
One of the most widely adopted models for DM haloes is the Navarro-Frenk-White (NFW) profile \citep{Navarro_1997}. Its functional form is given by
\begin{equation}
	\rho_{\text{NFW}}(r) =\frac{\rho_0}{\frac{r}{r_s}(1+\frac{r}{r_s})^2},
    \label{eq:NFW}
\end{equation}
where $\rho_0$ is the characteristic density and $r_s$ is the scale radius. This profile features an inner slope of $-1$ and an outer slope of $-3$, with a transition at the scale radius. It is common to parametrise NFW haloes in terms of concentration, $c = R_{200\mathrm{c}} / r_s$, and virial mass, $M_{200\mathrm{c}}$, instead of $\rho_{0}$ and $r_s$.

To generalise the definition of concentration across different functional density profiles we adopt the convention
\begin{equation*}
c = R_{200\mathrm{c}} / r_{-2}
\end{equation*}
where $r_{-2}$ is the radius where the density's logarithmic slope is equal to $-2$, i.e.,
\begin{equation}
	\frac{\text{d}\log\rho}{\text{d}\log r}\bigg\rvert_{r_{-2}} = -2.
	\label{eq:log_slope}
\end{equation}
For an NFW halo $r_{-2} = r_{\mathrm{s}}$.

To allow comparison in extrapolated halo mass across different DM profile models, we report $M_{200\mathrm{c}}$ for each model, following the dark-matter–rescaled definition of virial mass in hydrodynamical runs or the standard definition for DMO runs (see Section~\ref{sec:DM rescaling}). For profiles where a closed-form analytical expression for the mass enclosed profile can be derived, $M_{200\mathrm{c}}$ is computed analytically; otherwise it is obtained by numerically integrating the fitted density profile.

\subsubsection{Generalised NFW}
\label{sec:gNFW}
The generalised NFW (gNFW) profile\footnote{The phrase `generalised NFW' is not consistently used to refer to the same parametric profile within the literature. While it is common to mean a profile with a free inner and outer slope, we use this to refer to a profile with a fixed outer slope and free inner slope, consistent with recent works inferring the Milky Way's mass from rotation curve analysis \citep[e.g.][]{Ou_2023}.} extends the standard NFW model by introducing a variable inner slope, $\gamma$, which allows for greater flexibility in describing the inner structure of dark matter haloes; the outer slope is fixed to be $-3$ as in the NFW profile. This additional degree of freedom allows the profile to represent diverse radial density profiles, including cored haloes that depart from the canonical NFW cusp.

The gNFW profile is defined as
\begin{equation}
	\rho_{\text{gNFW}}(r) =\frac{\rho_0}{r^\gamma(r+r_s)^{3-\gamma}},
    \label{eq:gNFW}
\end{equation}
where $r_s$ is the scale radius (i.e., the slope transition scale), and $\rho_0$ is the characteristic density. For this profile there exists a simple relationship between the scale radius and the radius we use for the concentration, $r_{-2}$, given by:
\begin{equation}
r_{-2} = (2 - \gamma) r_s.
\label{eq:gNFW_slope_scale_degen}
\end{equation}
To ensure physically plausible fits, we constrain the slope parameter to the range $\gamma \in [0,2]$.

\subsubsection{Einasto}
\label{sec:Einasto}
The Einasto profile \citep{Einasto_1965} is another widely adopted model for describing dark matter halo density profiles, both in simulations and observational studies. Like gNFW, it allows for a cored or cuspy inner density, however, instead of a broken power law, it features a continuously varying logarithmic slope governed by an exponential function.

Its functional form is
\begin{equation}
	\rho_{\text{Ein}}(r) = \rho_{-2}\exp({-2/\alpha[(r/r_{-2})^\alpha -1]}),
    \label{eq:Einasto}
\end{equation}
where $\alpha$ is the shape parameter, $r_{-2}$ is the radius where the logarithmic slope is $-2$ (following Equation~(\ref{eq:log_slope})), and $\rho_{-2}$ is the density at $r_{-2}$. The parameter $\alpha$ controls the steepness of the profile, with larger values of $\alpha$ producing sharper central concentrations, and lower values yielding flatter or core-like distributions. As for gNFW, we ensure physical fits by restricting the shape parameter to $\alpha > 0$. For $\alpha \approx 0.16$ Einasto and NFW profiles look similar over radial ranges typically probed by simulations and observations.

\subsubsection{Contracted NFW}
\label{sec:Contracted_NFW}

Baryonic processes can significantly modify the inner dark matter distribution in Milky Way-mass haloes. Among several proposed parametrisations of these effects, we adopt the phenomenological model introduced by \citetalias{Cautun_2020}, which was calibrated using matched hydrodynamical and dark-matter-only simulations to relate the degree of halo contraction to the relative enclosed mass profiles of the two runs. Within this section we introduce the mathematical framework of the model and corresponding definitions, following the original authors' parameter conventions. Later, in Section~\ref{sec: Testing_Cautun+20_Model}, we explicitly test the model predictions against the ARTEMIS simulations, which were not used in the original development and calibration of the mass-ratio relation (Equation~\ref{eq:eta-chi}).

\citetalias{Cautun_2020} found that the ratio of enclosed dark matter mass,
\begin{equation}
\eta_{\mathrm{DM}}(r) = M_{\mathrm{DM}}(<r)/M_{\mathrm{DM, DMO}}(<r),
\end{equation}
correlates closely with the ratio of the \textit{total} enclosed mass,
\begin{equation}
\chi_{\mathrm{tot}}(r) = M_{\mathrm{DM, DMO}}(<r)/(M_{\rm DM}(<r)+M_{\rm bar}(<r)),
\end{equation}
between a DMO and hydrodynamical simulation of Milky Way--mass haloes. The relation is well described by a power law,
\begin{equation}
	\eta_{\mathrm{DM}} = A\chi_{\mathrm{tot}}^{B},
    \label{eq:eta-chi}
\end{equation}
with $A = 1.023 \pm 0.001$ and $B = -0.540 \pm 0.002$ calibrated from simulations.

The contracted profile can then be approximated as
\begin{equation}
	M_{\mathrm{DM, CNFW}}(<r) = M_{\mathrm{NFW}}(<r) f(\eta_{\mathrm{bar}}),
    \label{eq:Contracted_mass}
\end{equation}
with
\begin{equation*}
f(\eta_{\mathrm{bar}}) = [0.45 + 0.38(\eta_{\mathrm{bar}}+1.16)^{0.53}],
\end{equation*}
where $\eta_{\mathrm{bar}}$ is defined as the ratio between the baryonic mass and the NFW mass profile (or any halo model representing the DMO dark matter mass profile), scaled by the cosmic baryon fraction,
\begin{equation}
\eta_{\mathrm{bar}}(r) = \frac{1}{f_{\mathrm{bar}}}\frac{M_{\mathrm{bar}}(r)}{M_{\mathrm{DM, DMO}}(r)} .
\end{equation}

Equation~(\ref{eq:Contracted_mass}) implicitly assumes that the uncontracted halo follows an NFW distribution, which is why we refer to this as the contracted NFW (CNFW) model. In principle, other `contracted' models can be formulated equivalently by using another profile for the uncontracted halo model, provided it gives a good description of the halo's matter distribution in DMO simulations.

\section{The Effects of Baryonic Condensation on Dark Matter Haloes}
\label{sec:Contraction}

\begin{figure}
    \includegraphics[width=0.475\textwidth]{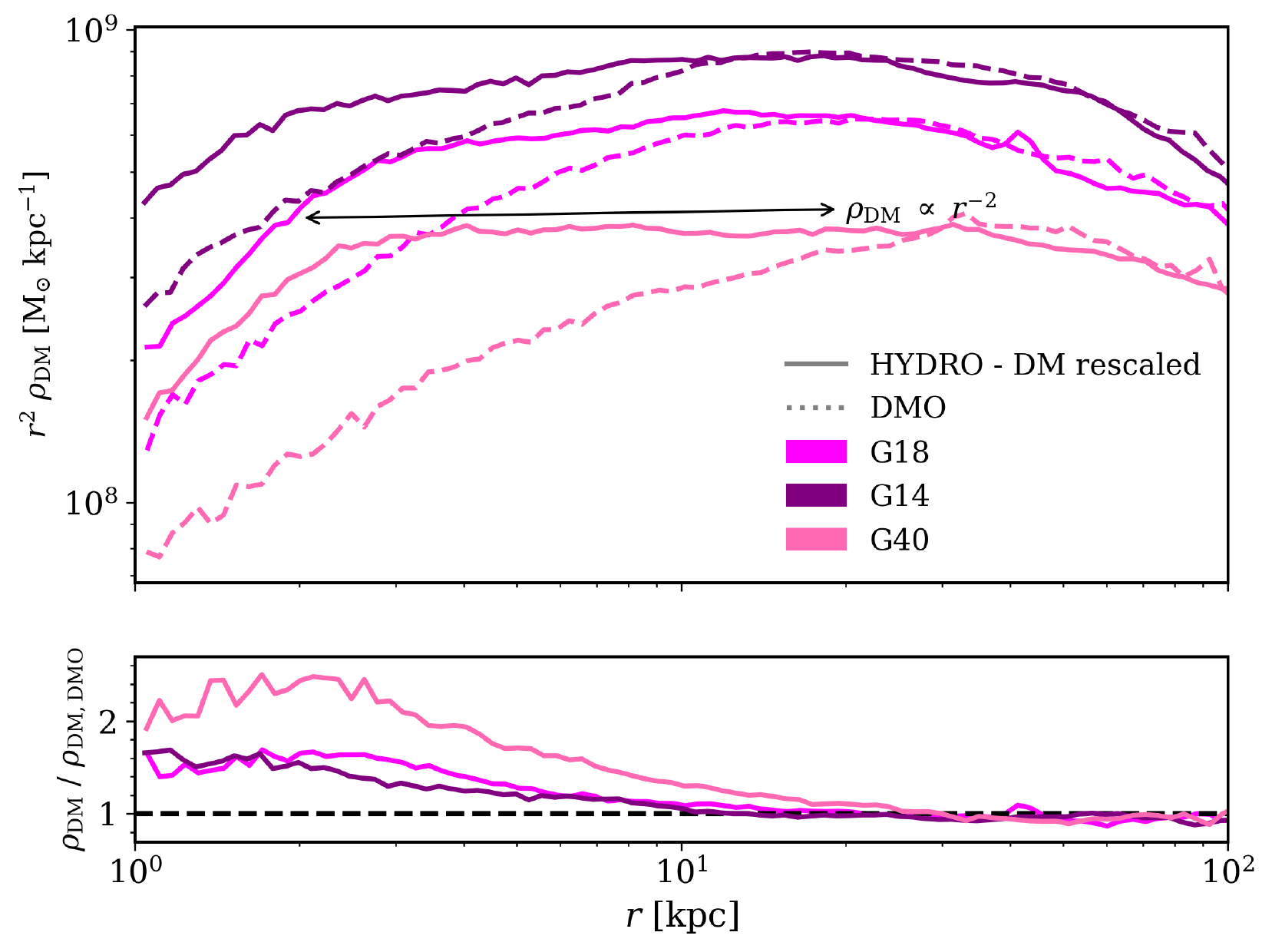}
	\caption{
    \textit{Top panel:} Spherically averaged dark matter density profiles for three representative ARTEMIS central haloes (G40, G18, and G14). Solid lines show the hydrodynamical profiles after rescaling the enclosed dark matter mass by the cosmic mean baryon fraction, while dashed lines of the same colours denote their dark-matter-only (DMO) counterparts.
    \textit{Bottom panel:} Ratio in dark matter density profiles between the rescaled hydrodynamical and DMO versions of the same haloes. The dashed black line indicates no difference between the two. 
    Colours in both panels distinguish the individual haloes, as indicated in the legend. The impact of baryons is most pronounced at radii $r \lesssim 20~\mathrm{kpc}$, where the hydrodynamical profiles deviate significantly from their corresponding DMO counterparts.
    }
	\label{fig:DM_rho_examples}
\end{figure}

We begin our analysis by examining the impact of baryonic physics on the inner structure of dark matter haloes. Specifically, we compare the spherically averaged DM density profiles of central haloes between the hydrodynamical and dark matter-only (DMO) runs of the ARTEMIS simulations. Throughout this work, we adopt the commonly used assumption of spherical symmetry, which is a reasonable approximation in the inner regions ($r \lesssim 20~\text{kpc}$), where baryonic processes tend to make the halo potential more isotropic \citep[e.g.][]{Frenk_88, Abadi_10, DiCintio_2014}. This region is also where rotation curve measurements are most constraining and where different theoretical halo models deviate more strongly from one another.

Perhaps the most striking consequence of baryonic physics in Milky Way–mass systems is the so-called contraction of the dark matter distribution, i.e. an enhancement in the central DM density compared to the DMO counterpart of a halo. This effect is illustrated in Fig.~\ref{fig:DM_rho_examples}, which shows the DM density profiles for three representative ARTEMIS haloes (G40, G18 and G14). Colours indicate which halo; hydrodynamical and DMO runs are distinguished by the line styles, solid and dashed, respectively.

Baryonic condensation enhances the central dark matter density typically for $r \lesssim 20~\text{kpc}$. The magnitude and location of this enhancement varies between haloes, reflecting differences in their baryonic content and assembly histories, as shown more directly by the density ratios in the bottom panel of Fig.~\ref{fig:DM_rho_examples}. For example, G40 (pink) exhibits the strongest response, reaching a maximum enhancement of $\rho_{\rm DM}/\rho_{\rm DM, ~DMO} \approx 2.5$ at $r \approx 2~\text{kpc}$, while the other haloes show more moderate peak increases of order $\sim 1.7$ at their respective peak radii. This baryon-induced contraction steepens the profile slope and brings it closer to an isothermal value. This is particularly evident for G40 again, where the hydrodynamical profile maintains a near-constant slope of -2 over $r \approx 2$--$20~\text{kpc}$. By contrast, its DMO counterpart only briefly approaches the isothermal value near $r \approx 20~\text{kpc}$, never sustaining it over a comparable interval. At larger radii, the baryon-induced enhancement in DM density typically diminishes and may even reverse, with the hydrodynamical profiles falling below their DMO counterparts. This outer suppression is substantially weaker in amplitude than the central enhancement (of order few percent rather than hundreds), but its magnitude also varies systematically between haloes, with more pronounced effects generally associated with differences in stellar mass. This behaviour, sometimes referred to as halo elongation, is visible for example in G18 and G40 beyond the slope transition scale. Around or beyond the virial radius ($r \sim R_{200\mathrm{c}}$), however, the hydrodynamical and DMO profiles converge in both enclosed mass and outer slope, suggesting that baryonic effects become negligible at sufficiently large scales; see also Fig.~\ref{fig:Rescaling}.

These findings are in good agreement with results from other large hydrodynamical simulation projects -- including IllustrisTNG, EAGLE, Auriga, and APOSTLE -- which similarly show that Milky Way-mass haloes are, on average, contracted \citep[e.g.][]{Schaller_15, Grand_2017, Pillepich_2017, Cautun_2020}. This contraction signature is not universal in amplitude or radial dependence, however: recent cross-suite comparisons \citep{Hussein2025, DREAMS2026} show that simulations such as Auriga, EAGLE and TNG where the ISM is modelled with an effective equation-of-state and the star formation density is relatively low are well described by simple adiabatic contraction (AC), whereas simulations which model strong, bursty feedback -- such as FIRE-2 -- show substantial differences, including a DM density enhancement stronger than what predicted by AC and, in some instances, even central core formation at Milky Way masses \citep{Lazar2020}. Differences in subgrid physics and modelling choices are therefore known to affect the DM distribution in hydrodynamical realisations of DM halos, in particular the modelling of bursty and stronger feedback in FIRE-like simulations is likely the source of disagreement on the signature of contraction. Given ARTEMIS shares a lot of modelling details with the suites \cite{Cautun_2020} calibrated against -- especially EAGLE\_recal -- we expect close agreement in the signature of contraction we observe and their model, nonetheless the recalibration of stellar feedback parameters which ARTEMIS involved makes the test which we present in the next section necessary to understand whether their contraction model can be applied to our halos.

\subsection{Modelling baryon-induced halo contraction}
\label{sec: Testing_Cautun+20_Model}

\begin{figure}
    \includegraphics[width=\columnwidth]{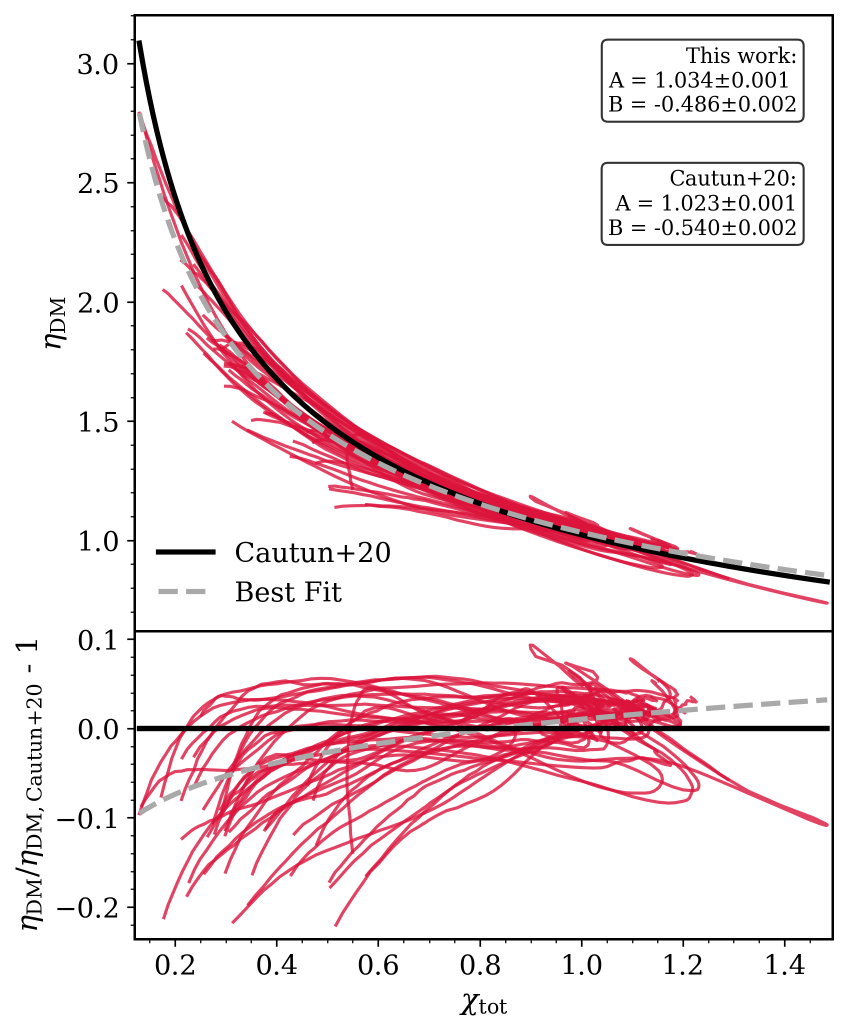}
	\caption{Relation between the enclosed dark matter mass ratio, $\eta_{\mathrm{DM}}$, and the enclosed total mass ratio, $\chi_{\mathrm{tot}}$, for all central ARTEMIS haloes. Ratios are computed from the spherically averaged cumulative mass profiles in the hydrodynamical and DMO versions of each halo.
	\textit{Top panel:} $\eta_{\mathrm{DM}}$–$\chi_{\mathrm{tot}}$ relation for individual haloes (dark red curves). The solid black line shows the \citetalias{Cautun_2020} best-fit power law, and the dashed grey line shows our best-fit to the ARTEMIS sample.
	\textit{Bottom panel:} Fractional deviation of each halo from the \citetalias{Cautun_2020} prediction. 
    The ARTEMIS haloes follow the same deterministic power law relation to within a few per cent, with both median offset and sample scatter within the \citetalias{Cautun_2020} model quoted bounds, confirming that the contraction formalism remains internally consistent when applied to a simulation suite with a similar (EAGLE-based) feedback implementation.}
	\label{fig:Cautun+20_ARTEMIS}
\end{figure}

A number of models have sought to describe baryon-induced contraction using adiabatic formalisms, in which the dark matter distribution in hydrodynamical simulations is assumed to conserve the same action integrals as in its corresponding DMO counterpart \citep[e.g.][]{Blumenthal1986, Gnedin_2004, Sellwood_2005, Callingham_2020, Hussein2025}. These approaches provide physically motivated and interpretable descriptions of halo response from first principles. However, their implementation in analytical profile fitting can require additional assumptions about the baryonic component and halo evolution that are not always directly constrained in observational contexts.

Here, we adopt the semi-empirical model presented in \citetalias{Cautun_2020} as a complementary approach. This model relates the enclosed dark matter and total mass profiles in hydrodynamical and DMO simulations through a deterministic power-law relation (see Sec.~\ref{sec:Contracted_NFW}). By directly coupling the local dark matter response to the enclosed baryonic mass profile, $M_{\rm bar}(<r)$, this framework accommodates a wide range of baryonic structures without requiring a prescribed analytical form or detailed knowledge of global quantities, such as the baryon fraction at the virial radius, which are often uncertain observationally. Given the feedback-dependence noted above, we treat this prescription as a practical tool suited to simulations with smooth stellar feedback, rather than as a universally valid model of contraction, and test its applicability to ARTEMIS directly below.

In this work, we apply this contraction prescription to the ARTEMIS haloes to assess how accurately it reproduces the dark matter structure in the hydrodynamical simulations. Once this is established, our goal is not to adjudicate on the detailed physics of contraction, but rather to quantify how the inclusion or neglect of baryon-induced contraction impacts the inferred dark matter distribution when fitting halo profiles to rotation curve data. The semi-empirical formalism provides a practical and flexible tool for this purpose. While the formalism can be generalised to other halo models, we adopt an NFW profile for the uncontracted halo due to its simplicity (two parameters) and its established success in describing DMO halo structure. We also emphasise that, unlike \citetalias{Cautun_2020} who developed this formalism to infer a mass model of the Milky Way from observational data, our aim is to benchmark its performance against commonly used analytic halo profiles that do not include contraction.

The \citetalias{Cautun_2020} power law -- i.e. Equation~(\ref{eq:eta-chi}) -- was calibrated using the Auriga, APOSTLE, and EAGLE\_recal suites, all of which model stellar feedback via an effective equation-of-state approach producing smooth, non-bursty outflow histories. ARTEMIS adopts the EAGLE subgrid physics model \citep{Schaye_2014}, with only the stellar feedback efficiency recalibrated relative to EAGLE\_recal \citep{ARTEMIS}, and so shares a similar feedback implementation. Close agreement with \citetalias{Cautun_2020} is therefore an expected outcome of this shared physics rather than evidence that the model holds across galaxy formation implementations generally -- a genuinely independent test would require comparisons with simulation suites with more significant differences in subgrid models. The comparison is nonetheless a necessary check: it confirms the prescription is internally consistent for the simulations analysed here, which is a prerequisite for using it as a quantitative tool in what follows.

To carry out this check, we compare the relationship between $\eta_{\mathrm{DM}}$ and $\chi_{\mathrm{tot}}$ for the ARTEMIS haloes against the original \citetalias{Cautun_2020} power law. Fig.~\ref{fig:Cautun+20_ARTEMIS} shows this relation, with individual halo profiles represented by dark red solid lines.

Following individual lines, that correspond to individual galaxies, highlights a subtle but noteworthy feature of the $\eta_{\text{DM}}$–$\chi_{\text{tot}}$ relation: haloes do not scatter randomly around the best-fit trend, but instead trace coherent trajectories. Most haloes begin in good agreement with the power-law at $\chi_{\text{tot}} \sim 1$ before systematically bending downward at smaller $\chi_{\text{tot}}$ values, which are associated with the innermost regions. This behaviour suggests that the contraction model performs best when averaged over halo populations, rather than when applied point-by-point within individual haloes.

Reassuringly, we find that ARTEMIS follows the same general trend as reported in the original study, with a best-fit power law that is remarkably close to the original: our inferred slope and normalisation differ by less than $10\%$ from the \citetalias{Cautun_2020} model. Furthermore, the bottom panel of Fig.~\ref{fig:Cautun+20_ARTEMIS}, which quantifies the fractional deviation from their best-fit, reveals that most haloes lie within the $\pm5\%$ band, and with the largest deviations in our sample remaining bounded by their quoted $20\%$ scatter envelope. The minimal systematic offset, and a halo-to-halo scatter within their quoted bounds, confirm that the power-law relation is internally consistent when applied to ARTEMIS -- as expected given a similar feedback prescription -- and validate its use as a quantitative tool for the rest of this work. It also strengthens the point that stellar feedback efficiency alone does not seem to have a first order effect on the response of DM haloes to the assembly of MW-like galaxies in their interiors, given the absence of significant differences between ARTEMIS and EAGLE\_recal.

\section{The Role of Contraction in Spatially Constrained Dark Matter Profile Fits}
\label{sec: Response of different profiles}

\begin{figure*}
	\centering
	\begin{tabular}{cc}
    	\includegraphics[width=0.48\textwidth, height=0.43\textheight]{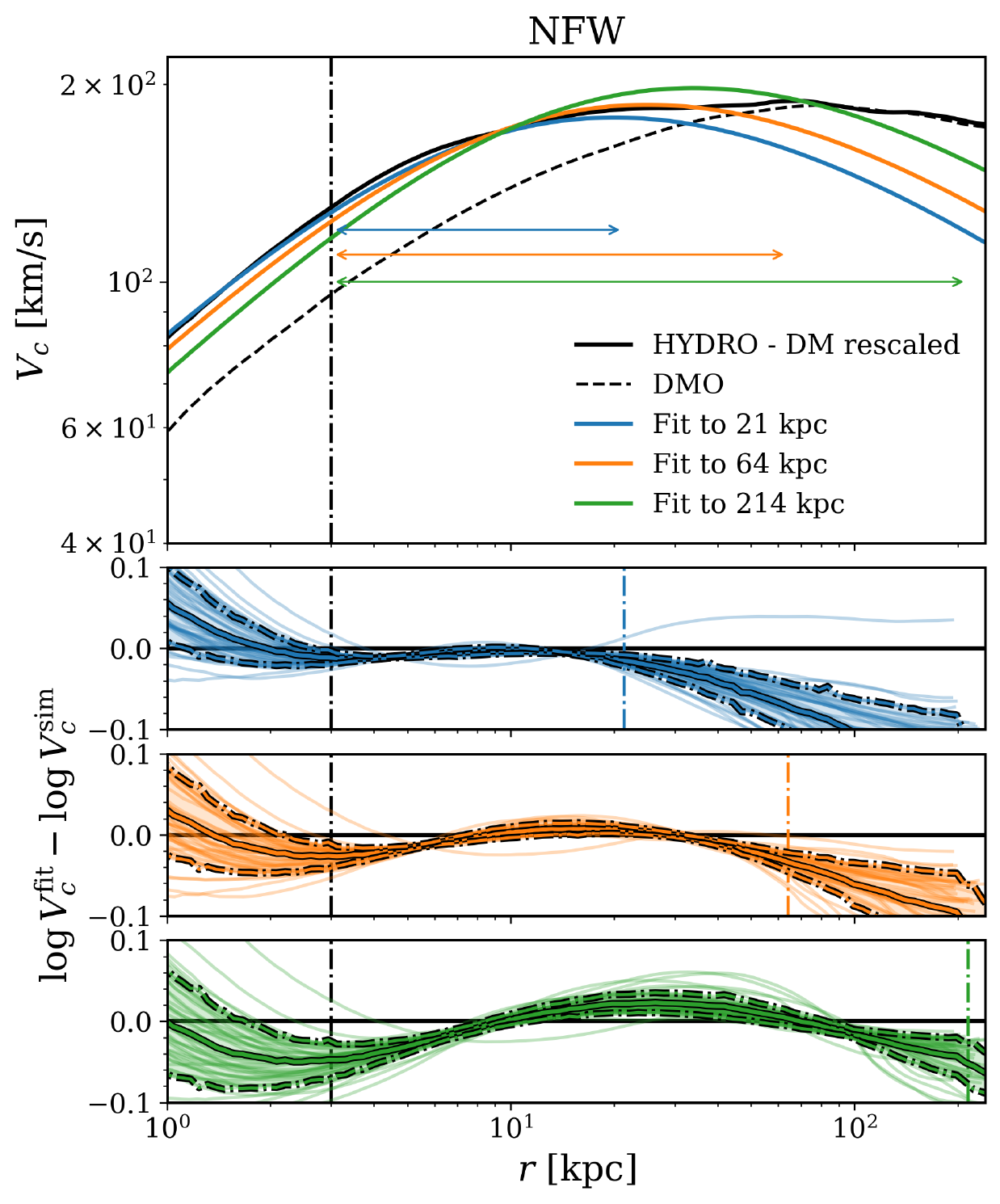} &
    	\includegraphics[width=0.48\textwidth, height=0.43\textheight]{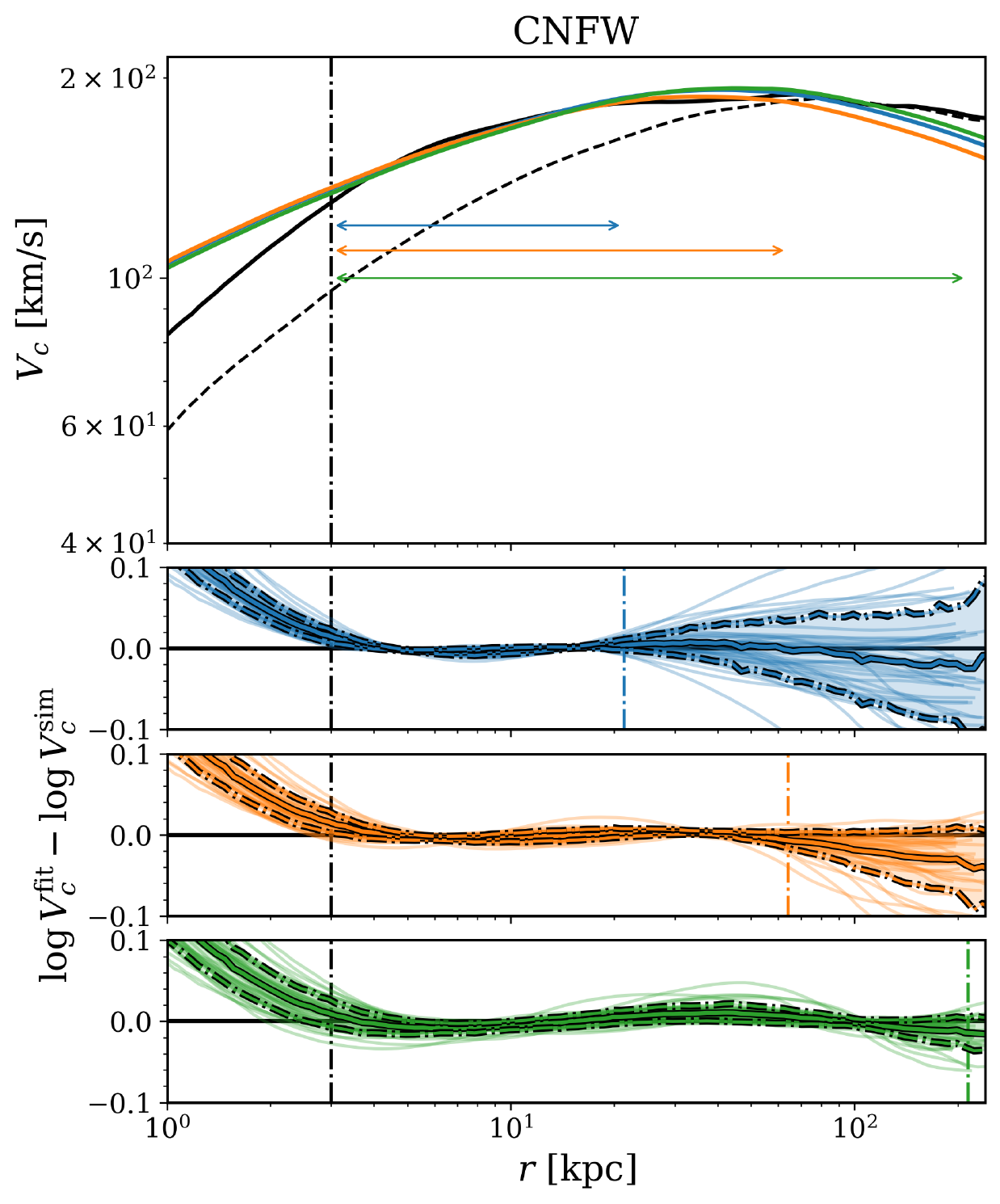} \\
    	\includegraphics[width=0.48\textwidth, height=0.43\textheight]{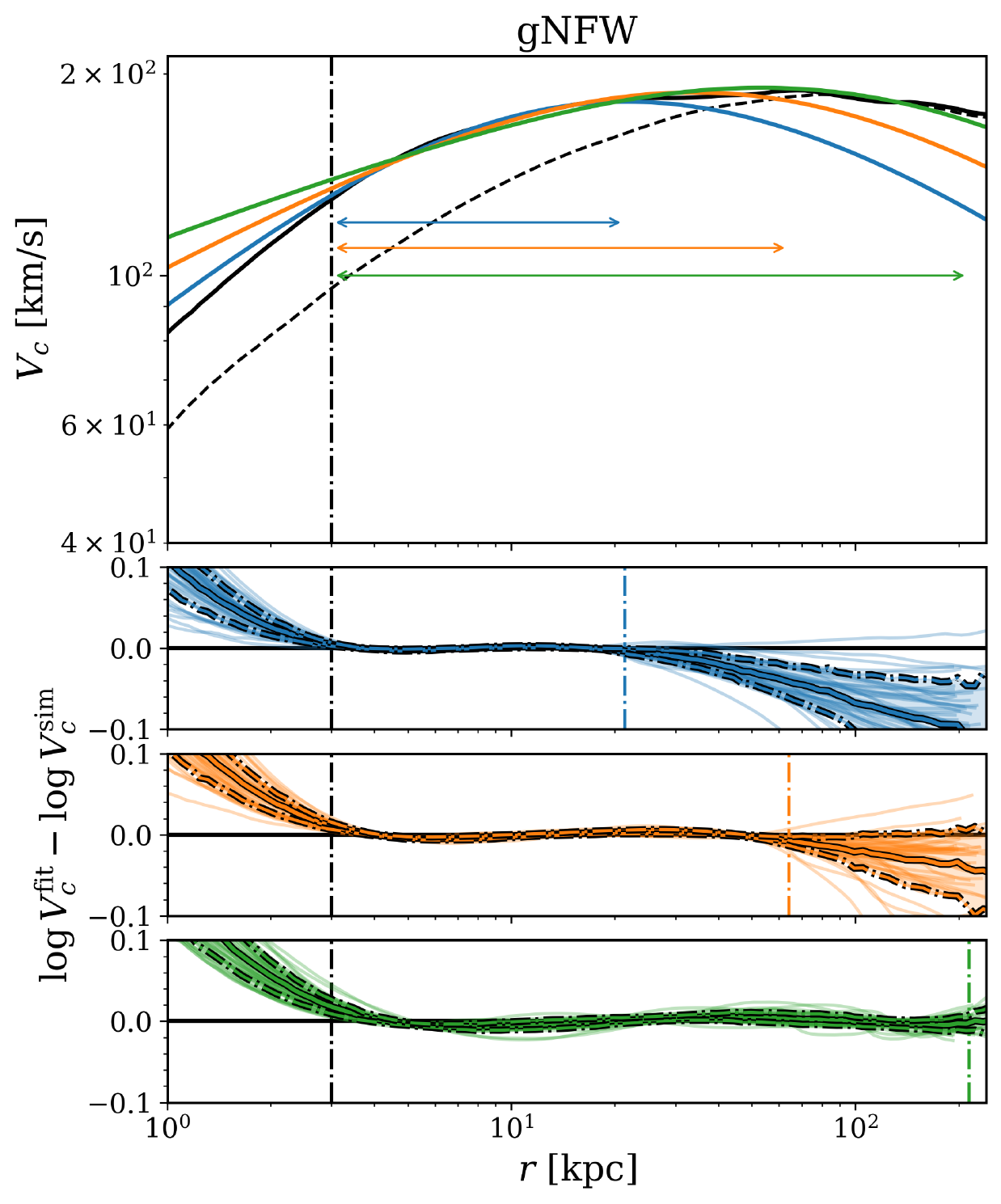} &
    	\includegraphics[width=0.48\textwidth, height=0.43\textheight]{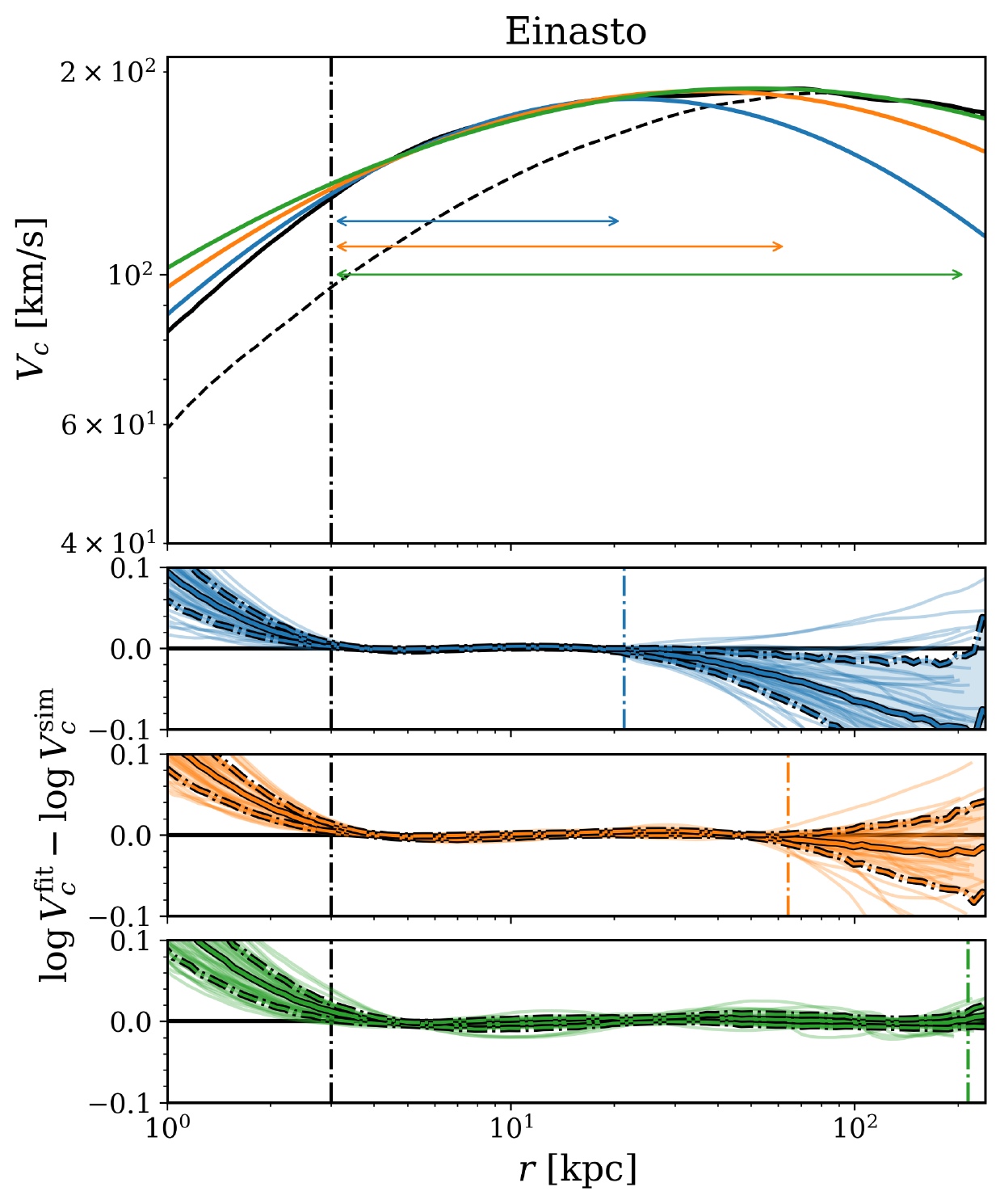} \\
	\end{tabular}
	\caption{Comparison of NFW, Contracted NFW (CNFW), gNFW, and Einasto fits when constrained to truncated radial ranges of simulated ARTEMIS haloes. Fits are applied to the DM circular velocity profiles of the DM-rescaled hydrodynamical runs.
	\textit{Top row of each panel:} Circular velocity profiles, $V_{c}(r)$, of a representative halo (solid black) and its DMO analogue (dashed black), together with best fits obtained when the outer fitting radius is limited to 21 kpc (blue), 64 kpc (orange), and 214 kpc (green). Coloured arrows indicate the radial interval used for each fit, following the same colour scheme.
	\textit{Bottom rows of each panel:} Logarithmic deviations (in dex) between the fitted, $V_c^{\rm fit}$, and simulated hydrodynamical, $V_c^{\rm sim}$, profiles. Thin curves show individual haloes, the thick curve the sample mean, and shaded regions the $1\sigma$ halo-to-halo scatter. The solid black line denotes zero deviation. In each sub-panel, the coloured vertical dash–dotted line marks the corresponding outer fitting radius.
	In all panels the vertical dash-dotted line marks the fixed inner fitting radius (i.e., $3~\mathrm{kpc}$).
	Across all profile families, restricted radial coverage increases systematic deviations in the extrapolated profile, with the magnitude and character of the bias strongly dependent on the chosen functional form.}
	\label{fig:Fitting_examples}
\end{figure*}

Having shown that baryons significantly affect the internal structure of DM haloes, particularly for radii $\lesssim 20$–$30~\mathrm{kpc}$, a natural question is how well can global halo properties be recovered from observations of the mass profile over only a limited radial range, especially one that largely overlaps with these contracted regions. This is particularly relevant for rotation curve analyses of the Milky Way and similar galaxies, which are most sensitive to the same regions where contraction is strongest. To examine the effects of restricted radial coverage, we fit halo models to the DM mass profiles over various radial ranges. While this approach does not try to reproduce a full mock observation pipeline, it does quantify how well the properties of the host DM halo can be measured assuming that the local DM potential can be accurately measured with minimal errors. In this sense, we are studying what is the fundamental limit to the information to be extracted when reconstructing the properties of the host DM halo using robust data but restricted to relatively small radii, and the potential systematics associated with this process.

We evaluate the performance of four widely used theoretical profiles -- Einasto, NFW, gNFW, and Contracted NFW (see Section~\ref{sec:fitting_pipeline}) -- to spatially restricted regions of our simulated haloes. Each fit region is defined by a fixed inner radius, $r_{\text{min. fit}}$, and a varying outer radius, $r_{\text{max. fit}}$, ranging from $10\%$ to $100\%$ of the median $R_{200\mathrm{c, DMO}}$ in our sample (i.e. $\bar R_{200\mathrm{c, DMO}} \approx 214 ~\rm kpc$). This strategy enables us to quantify how much radial coverage is required to reliably constrain bulk properties of the halo such as mass, $M_{200\mathrm{c}}$, and concentration, $c$.

\subsection{Fitting range sensitivity and profiles behaviour}
\label{sec: Fitting range sensitivity}

We begin by examining how well each profile captures the structure of DM haloes within the fitted radial domain. This step also serves as a sanity check: a model that fails to reproduce the halo structure within the fitted region is unlikely to yield meaningful extrapolation to larger radii. Figure~\ref{fig:Fitting_examples} presents a detailed view of profile behaviour across our model suite. Each panel corresponds to a different model -- NFW, Contracted NFW (CNFW), gNFW, and Einasto -- and compares the fits obtained by varying the outer fitting radius while keeping the inner radius fixed.

The top panel in each quadrant shows the DM circular velocity profile, $V_{\rm c}(r)$, of a typical DM-rescaled hydrodynamical halo (solid black) and its DMO counterpart (dashed black). We specifically show halo G2, with other haloes exhibiting qualitatively similar behaviour. Overplotted are best-fit parametric models constrained within three radial intervals: 3–21 kpc (blue), 3–64 kpc (orange), and 3–214 kpc (green). The innermost interval (3–21 kpc) corresponds to a typical range used in stellar rotation curve analysis. The vertical dash-dotted line marks the common inner boundary.

When comparing the blue (3-21 kpc fits) and green (3-214 kpc fits) curves, we notice a broadly consistent behaviour in all models which do not explicitly parametrise the effect of baryon-induced contraction (i.e. NFW, gNFW and Einasto). They all tend to shift their best fit parameters as more data is included -- reflecting the difficulty for these functional forms to model an extended radial range where the density is isothermal, and an outer halo largely untouched by baryonic processes (DMO-like). The CNFW model is less sensitive to this effect, as its formulation explicitly accounts for the role of baryons in shaping the local DM distribution. This results in less variance in halo parameters across different radial coverages, or, equivalently, stable fits. The behaviour just described is particularly noticeable in the example NFW profile (top left panel), where the 3-21 kpc fit (blue) gives a circular velocity peaking at $r \approx 10~\mathrm{kpc}$, whilst in the 3-214 kpc fit (green) it flattens at $r \approx 30~\mathrm{kpc}$ -- i.e. roughly a factor of 3 larger $r_{\rm peak}$.

The bottom three panels in each quadrant of Figure~\ref{fig:Fitting_examples} show the relative deviation between the best-fit and simulated circular velocity profiles for all haloes in our sample. We quantify this deviation as the difference in their logarithms, measured in dex -- i.e. $\log V_c^{\rm fit} - \log V_c^{\rm sim}$. The colours indicate the fitted radial range (as in the top panels), and each fitting range is displayed in a separate subplot. Thin lines represent individual haloes, while thick lines show the sample mean. Shaded regions denote the $1\sigma$ scatter.

The NFW fits reveal several clear trends. When only the inner galactic region is constrained, extrapolated circular velocities are systematically biased low -- by $\approx 0.10~\rm dex ~(26 \%)$ at $100~\rm kpc$ and $\approx 0.17~\rm dex ~(48 \%)$ near the virial radius -- despite minimal residual amplitudes and structure over the fitted range. This demonstrates that high formal fit quality within a limited radial interval (small compared to the virial radius) should not be mistaken for accurate inference of global halo properties. As seen in the bottom two panels of the first quadrant, widening the fitted radial range reduces the median offset at the virial radius, though the halo-to-halo scatter decreases more modestly. Notably, improving long-range accuracy comes at the cost of increased residual structure and larger scatter within the constrained domain, which is reflected in the enhanced wave-like patterns of the orange and green curves.

Interestingly, the more flexible three-parameter models -- gNFW and Einasto -- show similar qualitative behaviour. When fits are restricted to the inner galaxy (3–21 kpc, blue curves), they also yield substantially biased estimates of the circular velocity near the virial radius. Although the fits within the probed region are formally excellent, with absolute residuals never exceeding $0.005~\rm dex ~(1 \%)$ and virtually no scatter, the extrapolated circular velocity at the virial radius is still biased by $\approx -0.11~\rm dex ~(-22 \%)$ on average. As before, the scatter increases within the fitted domain and decreases in the extrapolation region with an increase in radial coverage, though sample variance is comparatively suppressed for the three-parameter models. Furthermore, while NFW fits never recover the circular velocity at the virial radius perfectly (median offset $\approx -0.05~\rm dex ~(-10 \%)$ even under the most favourable conditions), the gNFW and Einasto profiles can achieve negligible median bias -- but this requires unrealistically large radial ranges, spanning essentially the entire virialised region, to constrain their parameters.

As briefly mentioned when discussing the fit results from the representative halo shown in the top panel of each quadrant, but shown clearly for the entire ARTEMIS sample in the bottom three panels, the CNFW model yields broadly consistent fits almost independently of the radial coverage. When extrapolating fits with a range of order tens of kiloparsecs the median offset at the virial radius is $\approx -0.03~\rm dex ~(-7 \%)$, and $\approx -0.02~\rm dex ~(-5 \%)$ with a range of order hundred kiloparsecs. As for all models discussed, increasing the radial coverage leads to increased sample variance in the fitted domain and decreased variance in the extrapolated domain -- in this sense CNFW sits somewhere between the NFW and the three-parameter models (gNFW and Einasto). 

Interpreting these tendencies requires both mathematical and physical considerations. Mathematically, it is natural that parametric profiles adjust their degrees of freedom to improve the likelihood, resulting in formally higher quality fits within the probed radial domain. Physically, the preferred direction in parameter space is dictated by baryonic effects. As discussed in Section~\ref{sec:Contraction}, a prominent consequence of baryon-induced contraction is the emergence of an extended radial interval in which the density slope is very close to $-2$, forming a broad isothermal shell. As shown in Figure~\ref{fig:DM_rho_examples}, this region can span tens of kiloparsecs, typically bounded within $2$-$20 ~\rm kpc$, and features an enhanced density compared to a DMO realisation. Models that do not explicitly encode contraction physics can mimic this feature, but only by distorting the overall halo structure.  We note that the specific radial structure underlying the arguments that we present -- a broad, near-isothermal shell produced by smooth, centrally-concentrated contraction -- is itself a property of the ARTEMIS/EAGLE feedback model at MW-mass regimes, rather than a generic feature of baryonic physics in DM haloes. Simulations such as FIRE-2 with burstier star formation can show a qualitatively different radius-dependent response, including local core formation at the very innermost regions rather than monotonic contraction \citep{Lazar2020, Hussein2025}. The specific degeneracies we identify below, particularly the interplay between the scale radius and the width of the isothermal shell, are therefore not guaranteed to generalise to simulations with different feedback implementations, and may be weaker or absent where no comparably broad isothermal region is produced. We would nonetheless expect the qualitative conclusion -- that fits restricted to the inner, baryon-affected region distort global parameters -- to persist in some form: even in bursty-feedback simulations, baryonic effects on the DM profile remain on a similar magnitude and strongest near the centre and dilute outward, so profiles lacking an explicit contraction term would still have to compensate for this central enhancement at the expense of the fitted large-scale shape.

The NFW profile, being the least flexible among the models we consider, is most strongly affected. With no parameter governing the curvature of the profile, NFW must compromise both the scale radius ($r_s = r_{-2}$) and normalisation ($\rho_0$) to absorb the influence of the extended isothermal region. When the fitted radial range is small (3-21 kpc), sampling falls entirely within the degenerate isothermal region. To reproduce the observed density profile, NFW compresses $r_s$ inward toward $r \approx 2~\mathrm{kpc}$ (relative to the typical DMO value of $\sim 20~\mathrm{kpc}$), producing artificially high concentrations. The extrapolation to large radii with these distorted parameters then systematically underestimates circular velocities at the virial radius. As the radial coverage increases, the inferred $r_s$ shifts outward and partially approaches the true (DMO) value reflecting the higher statistical weight of outer regions, which are weakly affected by baryonic effects. However, because NFW profiles cannot accommodate the broadness of the isothermal shell, the fitted halos remain significantly over-concentrated and under-massive even with extensive radial information.

The more flexible gNFW and Einasto models achieve even better local fits by adjusting their curvature parameter ($\gamma$ and $\alpha$, respectively), but this extra flexibility simply allows them to overfit the isothermal shell's appearance in a narrow radial slice (compared to the virial radius). In both cases, parameters optimised for local residual minimisation carry little information about structure beyond the fitted domain, as the structure of the halo past $20$-$30 ~\rm kpc$ is largely uncorrelated with the isothermal shell. This results in behaviours similar to the NFW model, i.e. systematic biases in extrapolated circular velocities, as they all arise by the same tendency to optimise for a region that carries little information about the outer halo. In fact, only when the fitting range extends across nearly the full virialised region are the models' degrees of freedom sufficiently constrained to recover the true global structure; before then, strong degeneracies between the scale radius and inner slope parameters appear when examining the covariance matrices for gNFW fits (see Sec.~\ref{sec: Mass and concentration bias} for a more detailed discussion).

The CNFW model, on the other hand, incorporates explicitly the expected change to the DM structure due to baryon-induced contraction. Essentially, the model knows where to locate the isothermal shell given the observed baryonic mass profile $M_{\rm bar}(<r)$, thereby preventing the fitted parameters (e.g. $\rho_0$ and $r_s$) from compensating for baryon-induced contraction by distorting the global halo structure.

Breaking the degeneracy between fitted parameters and the physics of contraction results in stable fits across a wide range of radial scales for this model. This is evident in the representative halo shown, where the blue and green curves are nearly indistinguishable -- note that the former is constrained by an outer radius $1 ~\rm dex$ smaller than the latter.

These trends are not driven by numerical resolution or sampling choices; varying the number of radial bins in the analysis has a negligible effect on our results. Instead, the biases arise from the limited radial leverage of the fit, with profile parameters compensating for the locally enhanced density when baryonic effects are not modelled explicitly.

Degeneracies between $r_s$ and $\gamma$ for the gNFW model have already been highlighted when fitting the Milky Way’s rotation curve \citep[e.g.][]{Karukes_2019, Ou_2023}, indicating that the estimates of the Galactic DM halo structure might be affected by these systemic biases. Overall, our results demonstrate that high-fidelity fits -- i.e. formally small residuals -- within a limited radial range, particularly one that overlaps with the contracted region of a halo, do not necessarily guarantee an accurate recovery of global halo properties. While all models perform well where they are expected to -- in the regime where data are available -- their ability to correctly extrapolate beyond this region varies significantly, as we explore in greater detail in the next section.

\subsection{Model-dependent bias in extrapolated halo properties}
\label{sec: Mass and concentration bias}

\begin{figure*}
    \includegraphics[width=\textwidth]{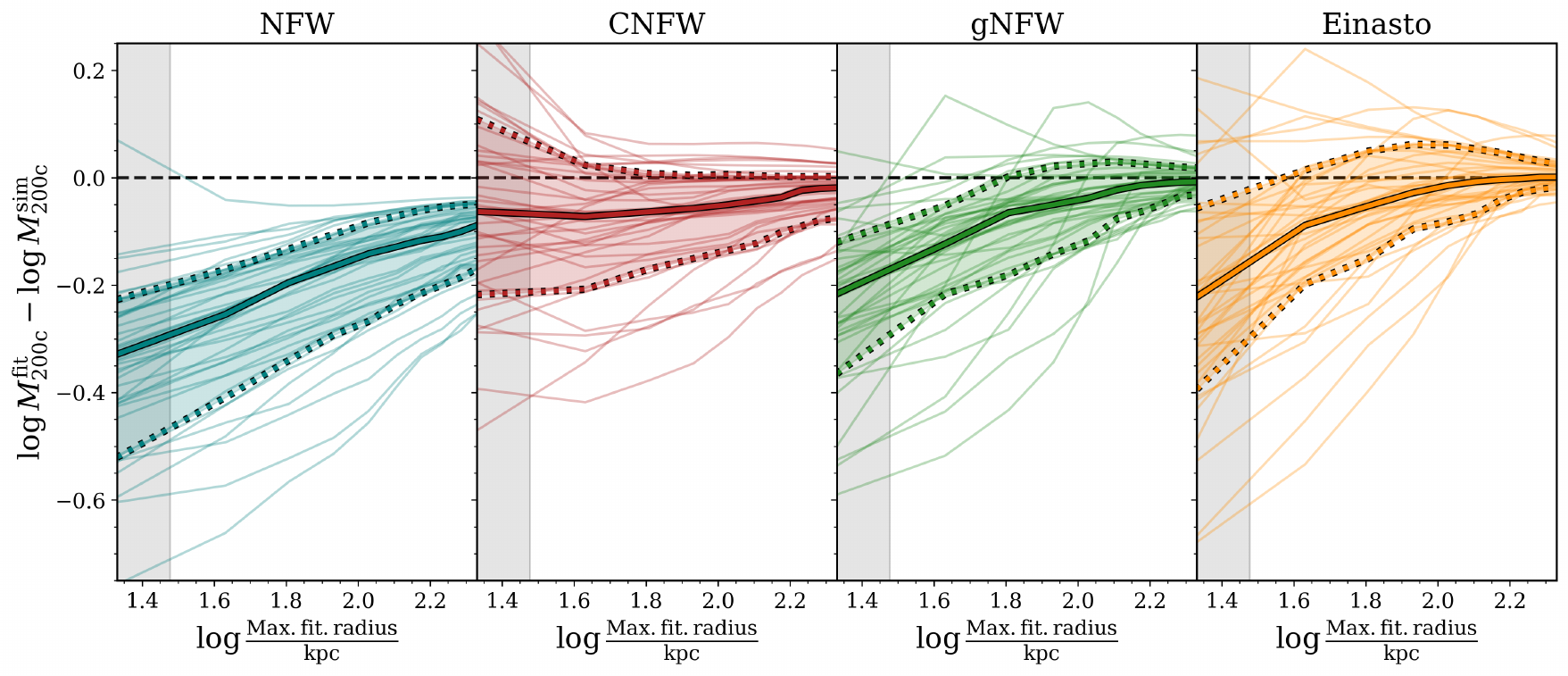}
	\includegraphics[width=\textwidth]{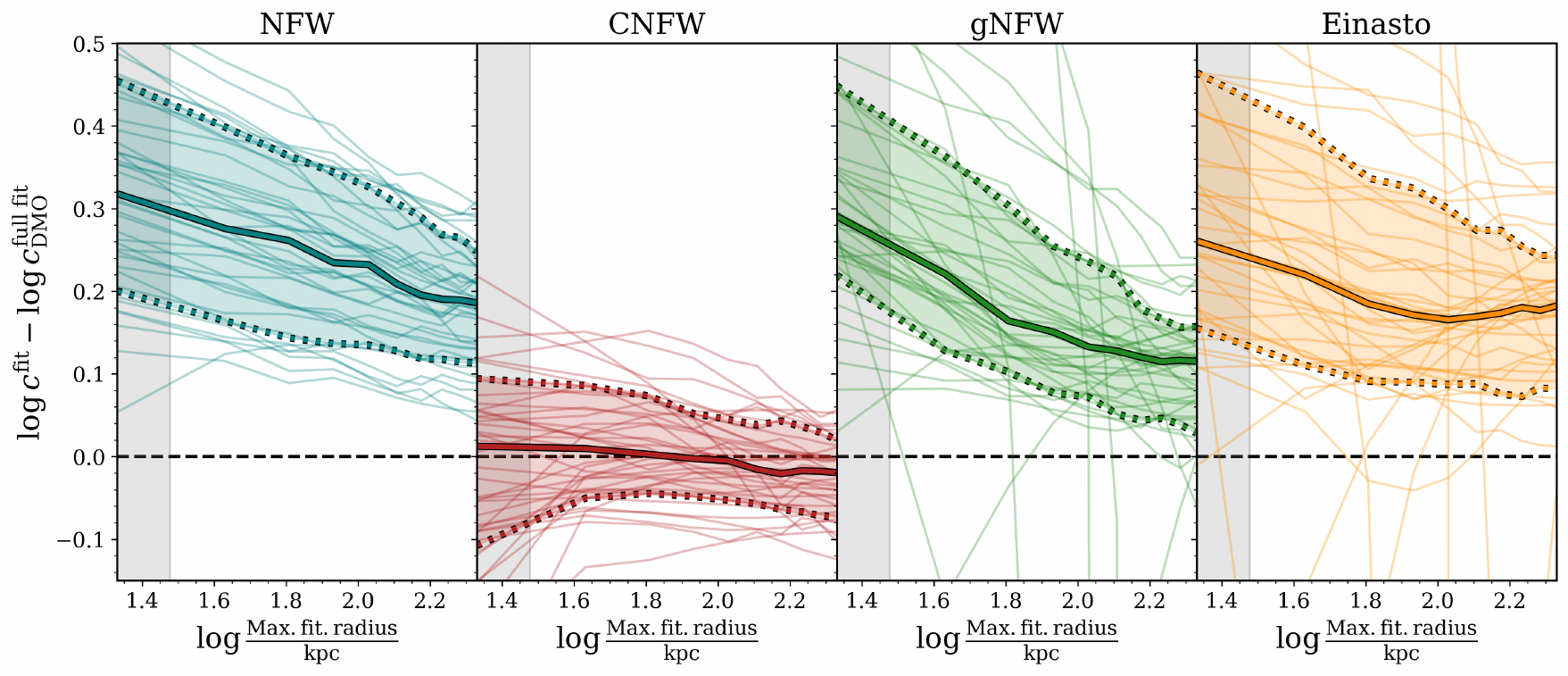}
	\caption{Model–dependent biases in extrapolated halo properties as a function of the maximum fitted radius, $r_{\rm max,fit}$. For each ARTEMIS halo we plot the logarithmic offset between the fitted parameter and its true/reference value.
	\textit{Top panel:} Logarithmic offset between virial masses -- $\log M_{200\mathrm{c}}^{\rm fit} - \log M_{200\mathrm{c}}^{\rm sim}$ -- against the logarithm of the maximum fitting radius.
	\textit{Bottom panel:} Same as the top panel but focusing on the concentration parameter. Hence showing deviations in the recovered concentration relative to that of the corresponding DMO halo -- $\log c^{\rm fit} - \log c^{\rm full~fit}_{\rm DMO}$.
	Thin coloured curves show individual haloes, thick curves the sample median, and shaded regions the $16^{\rm th}$–$84^{\rm th}$ percentile range. Columns, as well as colours, correspond to the four fitted models (NFW, CNFW, gNFW and Einasto, from left to right). In both panels, the grey shaded band indicates the typical observational range of the outermost radius probed by the Milky Way rotation curve. The figure demonstrates that restricting the fitted domain to small radii induces systematic, model-dependent biases in both halo mass and concentration, despite excellent local fits within the constrained region.}
	\label{fig:Biases on mass and concentration}
\end{figure*}

Having established that DM profile fits constrained to limited radial domains can, in some cases, exhibit significant residual structure in the extrapolated regions and signs of overfitting, we now focus explicitly on the impact of these effects on key structural properties. Specifically, we examine how the inferred values of halo mass, $M_{200\mathrm{c}}$, and concentration, $c$, vary with the choice of profile model and the radial extent used for its fit.  

Figure~\ref{fig:Biases on mass and concentration} presents a systematic evaluation of these estimates across the ARTEMIS halo sample, comparing best-fit values from profiles truncated at varying outer radii. Here different columns correspond to different models (NFW, CNFW, gNFW and Einasto; from left to right) and rows provide summary diagnostics for halo mass and concentration. Coloured thin lines indicate tracks from individual halos, whilst thick lines and shaded regions indicate the sample median and $68^{\rm th}$ percentiles (or sample scatter); the grey shaded band represents the realistic radial range currently probed by kinematic tracers in the Galaxy. Specifically, we quantify the bias in halo mass via the deviation between the logarithms of the inferred and values computed via the DM-rescaled $M_{200\mathrm{c}}$ convention (see Section~.\ref{sec:DM rescaling}) -- i.e., $\log M_{200\mathrm{c}}^{\rm fit} - \log M_{200\mathrm{c}}^{\rm sim}$. The bias in the inferred concentration is measured relative to the best-fit value when the full virialised region is used to constrain the parameters of the DMO counterpart of a halo -- i.e., $\log c^{\rm fit} - \log c_{\rm DMO}^{\rm full~ fit}$.

The NFW fits (leftmost column, light blue curves) lead to extrapolated values of $M_{200\mathrm{c}}$ systematically underestimating the true halo mass, with the $68^{\rm th}$ percentile of the population never intersecting the zero-deviation line across all choices of $r_{\rm max ~ fit}$. Even when fitting out to $r_{\rm max~fit} \approx \bar R_{200\mathrm{c, DMO}}$, the upper bound still falls short by $\approx 0.05 ~\rm dex ~(11 \%)$; conversely, the inferred concentrations are biased high -- with an offset of $\approx 0.12~\rm dex ~(32 \%)$ for the lower $16^{\rm th}$ percentile, as a direct consequence of the fitted scale radius, $r_s$, shifting inward to accommodate for the contraction of the halo. The halo-to-halo scatter in extrapolated mass decreases from a value of $\sigma_{\rm M{200}} \approx 0.15~\rm dex ~(41 \%)$ for fits limited to $21 ~\rm kpc$ to $\sigma_{\rm M{200}} \approx 0.07~\rm dex ~(17 \%)$ when limited to $214 ~\rm kpc$ -- roughly a factor of two over one decade in radius. The scatter in concentration estimates similarly declines by a factor of two at the typical virial radius from $\sigma_{c} \approx 0.13 ~\rm dex ~(35 \%)$ at $r_{\rm max ~ fit} = 21~\rm kpc$. From an observational perspective, these biases are particularly concerning. When only data within the inner $30~\rm kpc$ are available (denoted by the vertical shaded grey band), a regime common in Galactic and extragalactic rotation curve studies, the median extrapolated mass is underestimated and concentration is overestimated by a factor of two (i.e. $-0.3 ~\rm dex$ and $+0.3~\rm dex$, respectively). Moreover, in our most extreme case the halo mass is underestimated and concentration overestimated by nearly a factor of five. These results imply that NFW fits in the MW-mass regime are unreliable for precision estimates of $M_{200\mathrm{c}}$ and $c$, especially when extrapolating from parameters constrained using the inner, contracted galactic regions.

Looking at the inferred halo mass when assuming a gNFW or Einasto profiles (top row, last two columns; dark green and orange curves, respectively), they start with large yet comparatively suppressed underestimates of the halo mass, with median offsets of $-0.21 ~\rm dex ~(-38 \%)$ when the inner 21 kpc are probed, and reach negligible deviations when the maximum fitting radius is pushed to $\log r_{\rm max ~ fit}/\mathrm{kpc} \gtrsim 2$. Their halo-to-halo scatter differs slightly especially for $\log r_{\rm max ~ fit}/\mathrm{kpc} \lesssim 2$, as the sample spread in gNFW fits is of $\sigma_{\rm M{200}} \approx 0.13 ~\rm dex ~(35 \%)$, compared to $\sigma_{\rm M{200}} \approx 0.17 ~\rm dex~(48 \%)$ in Einasto fits. This causes the $86^{\rm th}$ percentile in Einasto fits to cross the zero-deviation line at $\log r/\mathrm{kpc} \approx 1.6$, roughly $0.2 ~\rm dex$ earlier than gNFW fits. These models are also qualitatively similar in the recovered concentrations (bottom row, last two columns). Einasto fits restricted to 21 kpc exhibit a median offset and halo-to-halo scatter in recovered halo concentration of $+0.26 ~\rm dex~(82\%)$ and $0.15 ~\rm dex~(41\%)$ respectively, while gNFW fits exhibit a median offset of $+0.29 ~\rm dex~(95 \%)$ and halo-to-halo scatter of $0.12 ~\rm dex~(32 \%)$. Even when the fitting range is increased to the full halo (i.e. 3-214 kpc), there persists significant biases in median offsets and sample spread in estimated concentrations; quantitatively, gNFW fits show a median offset of $+0.12 ~\rm dex~(32 \%)$ and halo-to-halo scatter of $0.07 ~\rm dex~(17 \%)$, and Einasto fits median offset of $+0.20 ~\rm dex~(58 \%)$ and halo-to-halo scatter of $0.09 ~\rm dex~(23 \%)$.

Despite an intuitive explanation for the trends exhibited by the vast majority of the sample, a small number of haloes show erratic evolution of their extrapolated parameters, particularly for the gNFW and Einasto models (last two columns of Figure~\ref{fig:Biases on mass and concentration}). For this reason we quote the $16^{\rm th}$--$84^{\rm th}$ percentile range rather than the standard deviation throughout, as the latter is disproportionately sensitive to these non-Gaussian tails and would misrepresent typical model performance. We find these outliers to be associated with significant substructure near the virial radius (e.g. a massive satellite or close companion), which distorts the spherically-averaged circular velocity profile in a way that cannot be captured by smooth parametric forms; the additional flexibility of the gNFW and Einasto models allows them to chase these localised features via extreme excursions in their shape parameters, at the cost of poorly constrained concentration estimates. We examine a representative case in detail in Appendix~\ref{sec:appendix_outliers}.

Strikingly, the Contracted NFW (CNFW) model largely mitigates the systematic biases seen in the rest of our model suite, even when the fit is restricted to the inner region ($\lesssim 20 ~\rm kpc$). Here, the median offset in estimated $M_{200\mathrm{c}}$ is of $-0.06~\rm dex ~(-13 \%)$ and the halo-to-halo scatter is of $0.15 ~\rm dex ~(41 \%)$, meaning a non-negligible portion of the sample returns an estimate within $\pm 25~\%$ of the true value. However, estimates do not converge to the zero-deviation line even when the entire virialised halo is used for their fits, retaining a median offset and scatter of $0.02 ~\rm dex ~(4.7 \%)$ and $0.03 ~\rm dex ~(7 \%)$, respectively. Similar to the Einasto fits, the median concentration estimate does not evolve as the radial range increases, but the halo-to-halo scatter does modestly decrease with larger radial coverage. Offsets are always of order $\pm 0.02 ~\rm dex ~ (< 5 \%)$, the sample scatter evolves from $\sigma_{c} \approx 0.10 ~\rm dex ~(26\%)$ for $r_{\rm max ~ fit} = 21 ~\rm kpc$ to $0.05 ~\rm dex~(12\%)$ at the maximal range. We also note that this estimate should not be interpreted as the concentration of the contracted halo. By explicitly accounting for baryon-induced contraction, the CNFW model is able to robustly infer concentration values that trace the intrinsic structural parameters of the halo, rather than those imprinted by baryonic condensation.

\subsection{Implications for DM inference in MW-mass systems}
\label{sec:Implications of contraction on MW}

The results presented above highlight several points of practical relevance for both modelling strategies and the interpretation of Milky Way mass estimates.

First, our analysis shows that the CNFW model provides the least biased inferences of global halo properties, $M_{200\mathrm{c}}$ and $c$, when the available kinematic information is restricted to the radial ranges where baryon-induced contraction dominates. In the MW mass regime this is precisely where most observational tracers reside \citep[e.g. \textit{APOGEE}, \textit{Gaia}, see the discussions in][]{Eilers_2019, Ou_2023}. Although more flexible three-parameter models (Einasto, gNFW) yield excellent local fits with smaller formal residuals, their ability to capture the correct global structure is compromised unless the fitted domain extends to large radii.

Expanding on this point, we find that fits restricted to $50 \lesssim r/\mathrm{kpc} \lesssim 214$ remove the significant systematic bias in halo mass: for all profiles the median offset less than $1\%$, with sample scatter of $\approx 10\%$. This is not simply an effect of including the virial radius, as fits over the narrower 50–70 kpc range exhibit nearly identical behaviour, with only slightly increased halo-to-halo scatter. Inferred concentrations, on the other hand, become largely unconstrained in these outer-halo fits, with any bias being small compared to the sample scatter (never smaller than $\sim 100 \%$ for any profile); this reflects the fact that both radial ranges exclude the typical scale radius ($r_s \sim 10$--$20\, \mathrm{kpc}$ for Milky Way–mass systems in our sample). These results reinforce our interpretation that the biases in halo mass arise primarily from the inner halo, where contraction alters the DM density profile: models adjust parameters to accommodate the near-isothermal shell (i.e., $\rho \propto r^{-2}$), whereas the largely DMO-like outer halo provides unbiased constraints when sampled. They also suggest that, ignoring other potential observation systematics, kinematic probes residing in the outer halo should be considered of higher fidelity when attempting a precision estimate of the Milky Way's halo mass.

Second, we find that internal diagnostics of fit quality are generally poor indicators of whether a given halo will yield unbiased extrapolations. Monte Carlo sampling of each best-fit parameters and covariance matrix (100 realizations per halo, per profile, per $r_{\rm max~fit}$) confirms that the intra-halo scatter is consistently one to two orders of magnitude smaller than the halo-to-halo scatter. Consequently, well-behaved fits -- i.e. small residuals, smooth profiles, stable parameter posteriors -- do not distinguish `good' from `bad' when extrapolating to larger radii.

Third, while the mean bias can be negligible there remains significant scatter between systems for all fitted profiles when using limited radii (typically $\approx 40$-$50\%$). In particular, internal goodness-of-fit metrics and the enclosed baryonic mass profile alone carry little information about the reliability of the extrapolation beyond the fitted radial range. The population scatter therefore represents a hard lower bound on the uncertainty in extrapolated halo properties -- even under the idealised assumptions adopted here, i.e. full knowledge of the baryonic distribution, negligible observational errors, and a perfectly reconstructed DM mass profile. In real applications the error budget can only be larger; hence, for fitting pipelines bounded by $r_{\rm max\,fit} \approx 20$–$30~\mathrm{kpc}$, we regard halo mass estimates with quoted uncertainties below $0.20~\mathrm{dex} ~(58\%)$ as likely over-optimistic in the Milky Way–mass regime.

Until recently, there was a broad consensus for the mass of the Milky Way ($\approx 10^{12} \mathrm{M_{\odot}}$), with many different tracers and methods agreeing within their quoted uncertainties \citep{Wang_20}. However, with an increase in the quality and quantity of data measurements using the inner galaxy have began to systemically infer lower halo masses, with significantly narrower quoted errors. \cite{Eilers_2019} inferred a mass of $M_{\mathrm{200c}} = (7.25 \pm 0.25) \times 10^{11} \, M_{\mathrm{\odot}}$ assuming and NFW DM profile. While the recent work of \cite{Ou_2023} infer mass of $M_{\mathrm{200c}} = (6.94 \pm 0.25) \times 10^{11} \, M_{\mathrm{\odot}}$ or $M_{\mathrm{200c}} = (1.81 \pm 0.25) \times 10^{11} \, M_{\mathrm{\odot}}$ when assuming a gNFW or Einasto DM profile, respectively. Our results show that when using profiles that do not account for halo contraction (i.e. NFW, gNFW and Einasto) it is common to underestimate the halo mass by a factor of $\approx 2$, and in the most extreme haloes $4$. This alone is enough to significantly reduce, and in most cases remove completely, the tension between these recent constraints on the Milky Way's virial mass and the traditional $\approx 10^{12} M_{\mathrm{\odot}}$, without even considering other systematic errors. 

A more recent revision by \citet{Ou_25} is an interesting external check in this context. Their updated NFW and gNFW fits yield Milky Way masses of $\approx 7$--$8 \times10^{11} M_{\mathrm{\odot}}$, closer to the canonical $\approx 10^{12} M_{\mathrm{\odot}}$ value than their Einasto fit ($\approx 10^{11} M_{\mathrm{\odot}}$), and both are also in better agreement with independent outer-halo probes (see their Fig.~10). As noted above, fits restricted to the outer halo show no systematic tendency to bias the recovered mass, since this region lies beyond the reach of contraction, making agreement with outer probes a stronger indicator of a reliable extrapolation than internal goodness-of-fit metrics, which carry little information about extrapolation accuracy (Section~\ref{sec: Fitting range sensitivity}). Within our sample, Einasto fits remain less biased than NFW on average, and recover systematically higher $M_{200c}$, for essentially any choice of maximum fitting radius. However, the halo-to-halo scatter in this bias is substantial: at a fitting radius comparable to that adopted by \citet{Ou_25} ($\sim 30$~kpc), haloes for which an NFW fit is less biased and extrapolates more reliably than an Einasto fit lie well within the $1$--$2\sigma$ range of the population. The fact that NFW and gNFW outperform Einasto for the Milky Way in \citet{Ou_25} is therefore consistent with our results: it reflects a plausible individual realisation of the halo-to-halo scatter we find, rather than a departure from the population-level trend.

We additionally examined the impact of the radial fitting range on the inferred DM density at the Solar radius, $\rho_{\rm DM}(r_\odot)$ where $r_\odot = 8 ~\rm kpc$. We find that all models except the standard NFW yield highly precise estimates: median deviations are $\lesssim 0.01~\mathrm{dex} ~ (2\%)$, and the halo-to-halo scatter never exceeds $0.02~\mathrm{dex} ~ (5\%)$. Only the NFW profile exhibits a noticeable bias, typically inferring densities too high by 5 \% (for 3-21 kpc fits) and up to 20\% for larger fitting ranges.
These results indicate that while global properties can be strongly biased when using restricted radial ranges, the local DM density at the Solar radius is remarkably robust to these modelling choices.

Taken together, these findings reinforce the view that halo contraction is a critical ingredient in dynamical modelling at Milky Way masses. They also provide a quantitative pathway for re-interpreting rotation-curve-based MW mass estimates in a manner consistent with the behaviour of realistic hydrodynamical galaxy formation simulations. Crucially, the biases we identify are not random; they follow systematic trends that depend on how contraction steepens the inner density profile and on how much of the halo is radially sampled. This implies that any inference relying solely on inner–halo kinematics necessarily encodes information about contraction, even if the model used does not. This perspective naturally motivates a broader question: to what extent do the biases introduced by different modelling choices distort the recovered concentration–mass relation itself? The next section examines this issue directly by comparing how NFW and Contracted NFW fits populate the $c$–$M_{200\rm c}$ plane as a function of radial coverage.

\subsection{The impact of modelling contraction on the inferred $c$–$M_{200\rm c}$ relation}
\label{sec: c-M relation}

\begin{figure*}
	\includegraphics[width=\textwidth]{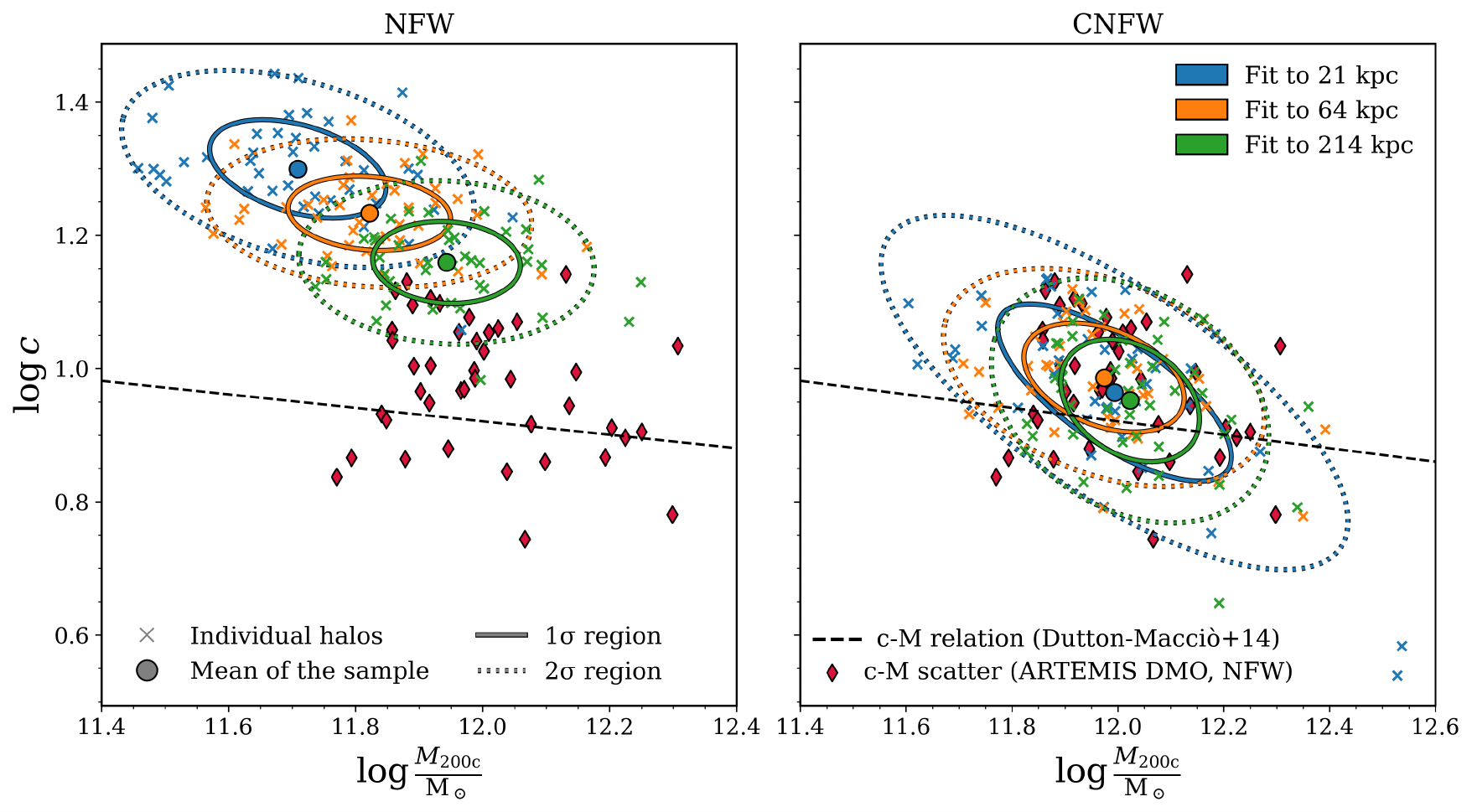}
	\caption{Recovered concentration–mass relations extrapolating from NFW (left panel) and Contracted NFW (right panel) fits. Colored crosses show individual haloes, with colors indicating the maximum fitted radius. Large filled circles give sample means for each fitting range, and solid/dashed ellipses indicate the corresponding 1$\sigma$/2$\sigma$ covariance contours.
    The black dashed line marks the $c$–$M_{200\rm c}$ relation from \citet{Dutton_2014}, shown as a reference. Dark red diamonds show the intrinsic $c$–$M_{200\rm c}$ distribution of the ARTEMIS DMO haloes, measured directly from the full profiles, and therefore provide the appropriate simulation-specific benchmark for comparison.
    The figure illustrates how, even with limited radial coverage, modelling baryon-induced contraction helps in recovering the correct underlying concentration–mass relation.}
	\label{fig:NFW-CNFW in c-M plane}
\end{figure*}

The concentration–mass relation is a cornerstone of $\Lambda$CDM structure formation, encoding the connection between halo mass, assembly history, and internal structure \citep[e.g.][]{Navarro_1997, Bullock_2001, Dutton_2014, Ludlow_2016}. In dark-matter–only (DMO) simulations this relation follows a tight, monotonic trend, with more massive haloes being less concentrated. In hydrodynamical simulations, however, baryonic processes -- including halo contraction, star formation, and energetic feedback -- can alter the inner density profiles and shift haloes away from the DMO prediction \citep[e.g.][]{Gnedin_2004, DiCintio_2014, Schaller_15}. Whether the canonical DMO relation remains a reliable benchmark for real galaxies, particularly at Milky Way masses, is therefore an important open question.

Earlier we showed that virial masses and concentrations inferred from NFW fits can be significantly biased when baryon-induced contraction is not modelled, with the magnitude and direction of the bias depending strongly on radial coverage. It is therefore natural to ask whether such mis-modelling affects the population-level $c$–$M_{200\rm c}$ relation itself, and whether a model that incorporates contraction (CNFW) can recover the expected scaling even when only the inner $\sim 20$ kpc are available. Figure~\ref{fig:NFW-CNFW in c-M plane} examines this by comparing how NFW and CNFW fits populate the $c$–$M_{200\rm c}$ plane for different choices of maximum fitted radius (left and right panels, respectively). Colored crosses show individual haloes, with colors indicating the maximum fitted radius -- blue for 3-21 kpc fits, orange for 3-64 kpc fits and green for 3-214 kpc fits (as in Figure~\ref{fig:Fitting_examples}). Large filled circles give sample means for each fitting range, solid/dashed ellipses indicate the corresponding 1$\sigma$/2$\sigma$ contours and the dashed black line represents the theoretical $c$–$M_{200\rm c}$ relation from \citet{Dutton_2014}, also referred to in the figure as Dutton-Macciò+14. We emphasise that the ARTEMIS DMO haloes themselves provide the most direct benchmark for assessing the accuracy of the recovered $c$–$M_{200\rm c}$ relation. These are shown in Fig.~\ref{fig:NFW-CNFW in c-M plane} as dark red diamonds, constructed from fits to the full radial extent of the corresponding DMO simulations. The \citet{Dutton_2014} relation is included for reference only, to provide a commonly used external comparison. In practice and as shown in the figure, ARTEMIS haloes scatter around this relation with a small systematic offset, consistent with differences in cosmology and halo selection, and alternative calibrations (e.g. a concentartion-mass relation from \citealt{Ludlow_2016}) would shift the line vertically by only $\sim 0.02$ dex in this mass range, well below the effects discussed here.

As shown in the left panel of Fig.~\ref{fig:NFW-CNFW in c-M plane}, standard NFW fits systematically overestimate concentrations while underestimating halo masses for all choices of radial coverage. The discrepancy is largest when the fit is restricted to $r \lesssim 20$ kpc, where haloes lie well away from the DMO relation by more than $2~\sigma$, even though the individual mass and concentration offsets fall within this range (see Fig.~\ref{fig:Biases on mass and concentration}). Crucially, the joint bias does not move haloes parallel to the standard $c$–$M_{200\rm c}$ relation. Instead, the correlated increase in concentration and decrease in mass shifts haloes roughly perpendicular to the expected trend, displacing them systematically off the canonical scaling. In other words, the biases reinforce one another in a direction that increases the apparent tension with the DMO prediction. This behaviour persists even for full-halo NFW fits, indicating that (i) the NFW functional form cannot reproduce the contracted profile, and (ii) the resulting parameter covariance drives coherent departures from the expected relation even when information well beyond the contracted region is included.

In contrast, CNFW fits (right panel of Fig.~\ref{fig:NFW-CNFW in c-M plane}) remain closely aligned with the DMO $c$–$M_{200\rm c}$ relation for all radial ranges, avoiding the systematic displacement seen for NFW fits. Even when restricted to 21 kpc, the sample mean lies on the expected scaling and the scatter remains small -- i.e., the median offset and 1$\sigma$ scatter mostly land on the dashed black line. Extending the fits to larger radii reduces the halo-to-halo scatter primarily along the mass axis, reflecting the increased radial leverage, but does not alter the central trend or the covariance orientation. An equivalent analysis for the DMO counterparts shows that mild systematic offsets persist for NFW fits even with full coverage and are qualitatively consistent with what shown in the right panel of Fig.~\ref{fig:NFW-CNFW in c-M plane}. These residual offsets are consistent with the intrinsic scatter and mild systematic offset of the ARTEMIS DMO haloes relative to the \citet{Dutton_2014} relation (see dark red diamonds in Fig.~\ref{fig:NFW-CNFW in c-M plane}), rather than indicating a shortcoming of the contraction model itself. Crucially, independent of the radial range used for the profile fitting, the posterior distribution of CNFW fits closely follows the intrinsic ARTEMIS DMO scatter in the $c$–$M_{200\mathrm{c}}$ plane, which constitutes the appropriate simulation-based benchmark. In contrast, NFW posteriors remain largely off the concentration-mass relation, whether compared to the \citet{Dutton_2014} relation or to the DMO haloes themselves.

It is worth clarifying the conceptual status of this comparison, since DM haloes in hydrodynamical simulations are not, in general, expected to reproduce the $c$--$M_{200\rm c}$ relation of DMO simulations. Baryonic condensation, cooling and feedback introduce non-gravitational physics with no counterpart in a DMO run, so there is no first-principles reason for a hydrodynamical halo's dark matter distribution to sit on, or near, the same locus in the $c$--$M_{200\rm c}$ plane as its DMO analogue. The relevance of the contraction formalism is not that it enforces such an expectation, but that it provides a physical route between the two: contraction arguments, whether adiabatic \citep[e.g.][]{Gnedin_2004} or semi-empirical \citep[e.g.][]{Cautun_2020}, relate the contracted profile to an assumed uncontracted, DMO-like progenitor through the (observable) enclosed baryon mass, via the local conservation of orbit-averaged action-like quantities or deterministic relations between baryon mass and total mass. The CNFW model implements this explicitly, parametrising the departure from an uncontracted NFW profile as a function of the enclosed baryon mass fraction, rather than assuming the hydrodynamical halo itself is DMO-like. Fig.~\ref{fig:NFW-CNFW in c-M plane} should therefore be read not as a test of whether hydrodynamical haloes obey the DMO $c$--$M_{200\rm c}$ relation but as a test of whether the uncontracted component recovered by the fit, once the baryon-induced distortion has been modelled and effectively subtracted, is consistent with the relation obeyed by the true DMO realisation of the same haloes. That CNFW recovers this relation, while NFW, gNFW and Einasto -- which do not separate a contraction term and instead absorb the baryonic distortion into a single set of DMO-motivated shape parameters -- systematically fail to, constitutes a non-trivial validation of the contraction ansatz itself: it shows that the formalism correctly isolates the baryon-induced response, leaving an inferred uncontracted profile whose statistical properties match those of the halo's actual DM-only counterpart. This is a property of the fitting model, not a prior assumption imposed on the hydrodynamical haloes.

We do not show gNFW and Einasto fits in Fig.~\ref{fig:NFW-CNFW in c-M plane}; however, we have verified their behaviour with analogous plots, which are presented in Appendix~\ref{app:gnfw_einasto_cm}. Qualitatively, both gNFW and Einasto follow trends similar to standard NFW. Fits restricted to small radii systematically overestimate concentrations while underestimating masses, displacing haloes away from the expected $c$–$M_{200\rm c}$ relation by significantly more than the $2\sigma$ population scatter. The main difference with NFW is that when the outer fitting radius exceeds $\sim 60~\mathrm{kpc}$, the $2\sigma$ covariance contours begin to overlap the expected relation, and when the full halo is sampled (up to 214 kpc) the $1\sigma$ ellipses also cross the canonical scaling. These results confirm that the coherent biases seen for NFW are broadly representative of profiles not explicitly modelling contraction, especially when restricted to the inner regions. 

These results have direct implications for analyses that impose the DMO $c$–$M_{200\rm c}$ relation as a prior when fitting halo models to observational data. Because NFW fits produce correlated biases -- overestimated concentrations associated with underestimated masses -- these shifts occur in a direction that does not follow the slope of $c$–$M_{200\rm c}$ relation itself. A tight prior based on the DMO relation therefore does not correct the joint bias; instead, it can result in apparently precise but systematically biased estimates, particularly when only the inner regions of the halo are probed, as is typical for the Milky Way. In contrast, CNFW fits recover the correct scaling without requiring a $c$–$M_{200\rm c}$ prior; the use of any likely becomes statistically innocuous rather than compensatory. The effectiveness of $c$–$M_{200\rm c}$–prior–based inference therefore depends primarily on the physical adequacy of the underlying halo model: when the profile family cannot reproduce and/or account for the contracted shape, no reasonable prior can eliminate the resulting joint bias.

In summary, explicitly modelling baryon-induced contraction is essential for recovering an unbiased concentration–mass relation from hydrodynamical haloes, especially when observational constraints reach only the inner $\sim 10~\%$ of the virial radius. CNFW fits preserve the population-level scaling across all radial ranges tested, whereas NFW fits introduce large, coherent distortions that propagates directly into analyses that employ $c$–$M_{200\rm c}$ priors (and similarly for gNFW and Einasto fits restricted to the inner regions). Importantly, these conclusions hold when comparing directly to the intrinsic $c$–$M_{200\rm c}$ distribution of the ARTEMIS DMO haloes, and are therefore not contingent on the choice of any external calibration of the relation. This underscores the need for contraction-aware models in dynamical studies of the Milky Way halo, even more so in efforts to interpret the concentration–mass relation in real galaxies. 

\section{Conclusion}
\label{sec:Conclusion}

In this work, we have quantified how baryonic processes modulate the structure of dark matter haloes in the ARTEMIS simulation suite, and how these effects propagate into the inference of global halo properties when observations are limited to the inner, baryon-dominated regions. Our results demonstrate that the inclusion of baryons produces an extended isothermal-like ($\rho \propto r^{-2}$) shell in the DM halo spanning $\approx 2$–$20~\mathrm{kpc}$, accompanied by an enhancement of the inner density (Fig~\ref{fig:DM_rho_examples}). This behaviour is consistent with previous findings from other hydrodynamical simulations and confirms that contraction is a generic outcome of galaxy formation in haloes of Milky Way mass.

We verified that the semi-empirical contraction model of \citet{Cautun_2020} accurately captures this response within ARTEMIS. The ARTEMIS haloes reproduce the same deterministic power law relation between enclosed dark matter and total mass ratios in hydrodynamical against DMO runs, with deviations from this trend remaining within the originally quoted scatter (Fig.~\ref{fig:Cautun+20_ARTEMIS}). This agreement reinforces the physical interpretation of contraction as a predictable coupling between the baryonic and dark matter components.

A central goal of this study was to assess how the limited radial coverage available to kinematic tracers and the choice of halo model impacts the recovery of structural halo parameters. By fitting several widely used parametric profile -- NFW, gNFW, Einasto, and CNFW -- to restricted radial ranges, we found:
\begin{enumerate}
	\item Excellent local fits do not guarantee accurate extrapolation. When constrained only by radii $\lesssim 20$–$30~\mathrm{kpc}$, NFW, gNFW, and Einasto models infer systematically biased halo mass and concentrations, even when their residuals within the fitted domain are minimal. Moreover, internal metrics -- such as parameter degeneracies or posteriors -- often carry little to no information on the quality of the extrapolation.
	\item The biases arise due to baryon-induced contraction creating a broad isothermal shell that strongly influences local curvature but carries little information about the outer halo. Models without explicit contraction physics compensate by distorting their fitted parameters, typically producing over-concentrated and under-massive haloes.
	\item The Contracted NFW (CNFW) model -- which incorporates baryonic effects through the \citet{Cautun_2020} formalism -- exhibits substantially more stable behaviour. Even when fitted only to $\sim 20~\mathrm{kpc}$, CNFW returns extrapolated halo masses within $\sim 0.06$ dex (15\%) of the true value on average, with a significant fraction of haloes accurate to within $25\%$. While some scatter remains, CNFW is the only model that avoids the strong systematic biases observed for the NFW, gNFW, and Einasto profiles. Importantly, it yields concentration values that trace the intrinsic structural parameters of the primordial halo, rather than those imprinted by baryonic condensation. Furthermore, it is the only model in our suite able to recover the underlying DMO concentration-mass relation when only the inner regions are probed.
	\item Increasing the maximum fitting radius toward the typical virial radius in our sample ($\bar R_{200, \rm DMO} \approx 214~\rm kpc$) substantially improves the recovery of halo mass and, to a lesser extent, concentration for all profile families — NFW, gNFW, Einasto, and CNFW. The improvement is smallest for CNFW, which already achieves very stable fits through explicit contraction modelling. It is worth noting that NFW estimates never fully reach zero median deviation in halo mass, and that all non-contraction-aware models continue to exhibit significant bias in concentration even when the full halo is probed. In practice, we find that accurate (1\% bias using CNFW, gNFW and Einasto) and precise (10\% scatter using CNFW, gNFW and Einasto) recovery of halo mass is best achieved using kinematic tracers restricted to the outer halo, largely unaffected by contraction (i.e., $r \gtrsim 50~\rm kpc$), whereas inner-halo coverage is more informative for constraining concentration provided a contraction-aware model.
\end{enumerate}

Taken together, our findings have direct implications for dynamical studies of the Milky Way and galaxies of similar mass. Kinematic data for the inner Galaxy primarily probe regions shaped by contraction; hence, widely used DM profiles can yield biased halo mass and concentration estimates even in the absence of observational uncertainties. This suggests that discrepancies among recent Milky Way mass estimates may originate not only from data choices or modelling assumptions, but also from the intrinsic limitations of profile forms that do not encode baryonic physics. Our results therefore advocate for the routine incorporation of contraction-aware models when interpreting mass tracers confined to the inner $\sim 30~\mathrm{kpc}$.

We emphasise that the `best' profile to use depends on the scientific question and application. For inferring the host halo properties, such as mass and concentration, the CNFW is the most reliable. While the flexible gNFW and Einasto profiles do not accurately extrapolate to larger radii, they do provide a better fit to the inner DM profile where we have the best kinematic data. By providing a more accurate description of the inner DM halo, these profiles are likely an equivalent or better choice for direct DM detection studies or searches for DM annihilation signals in the Galactic centre.

Future work will benefit from extending this analysis to a large observational rotation curve compilation applying contraction-calibrated models directly to MW-mass systems, especially given the (expected) reliable extraction of the underlying concentration-mass relation plane without the need of imposing the theoretical relation as a prior. Nonetheless, the conclusions from the present study are clear: baryon-induced contraction is both universal and dynamically significant, and accounting for it is essential for achieving unbiased inferences of the global properties of Milky Way–mass dark matter haloes.

\section*{Software}

The following software packages were used in this work:
\textsc{numpy} \citep{Numpy},
\textsc{scipy}\footnote{docs.scipy.org/doc/},
\textsc{pandas} \citep{pandas},
\textsc{astropy} \citep{Astropy},
\textsc{h5py} \citep{h5py}.

\section*{Acknowledgements}
DD acknowledges support by the Royal Society through a Dorothy Hodgkin Fellowship (DHF/R1/231105).
S.T.B. and A.F. acknowledge support by a UK Research and Innovation (UKRI) Future Leaders Fellowship (grant no MR/T042362/1) and Sweden's Wallenberg Academy Fellowship.
ASF acknowledges support from UKRI grant ST/W006766/1.
IGM acknowledges support from STFC grant ST/Y002733/1.

This work used the DiRAC@Durham facility managed by the Institute for Computational Cosmology on behalf of the STFC DiRAC HPC Facility (www.dirac.ac.uk). The equipment was funded by BEIS capital funding via STFC capital grants ST/K00042X/1, ST/P002293/1, ST/R002371/1 and ST/S002502/1, Durham University and STFC operations grant ST/R000832/1. DiRAC is part of the National e-Infrastructure. This work has made use of NASA’s Astrophysics Data System Bibliographic Services.

For the purpose of open access, the author has applied a Creative Commons Attribution (CC BY) licence to any Author Accepted Manuscript version arising from this submission.

Author contributions: DD led the analysis, interpretation and presentation of all results under the supervision of SB and AF. Other authors contributed to creating the ARTEMIS simulations, and provided input on the interpretation of results and draft versions of the manuscript.

\section*{Data Availability}

The data underlying the plots in this article are available from the corresponding author upon request. The raw data from the ARTEMIS simulation suite are not publicly available; researchers wishing to access the simulations may contact the ARTEMIS collaboration.

\bibliographystyle{mnras}
\bibliography{references}

\appendix 
\section{The origin of extreme outliers in extrapolated halo properties}
\label{sec:appendix_outliers}

As noted in Section~\ref{sec: Mass and concentration bias}, a small number of haloes in our sample show erratic evolution of their extrapolated halo parameters as the fitted radial range is increased, particularly for the gNFW and Einasto models. Here we examine one such case in detail and contrast it with a representative halo, to clarify the origin of this behaviour.

\begin{figure*}
	\centering
	\includegraphics[width=\textwidth]{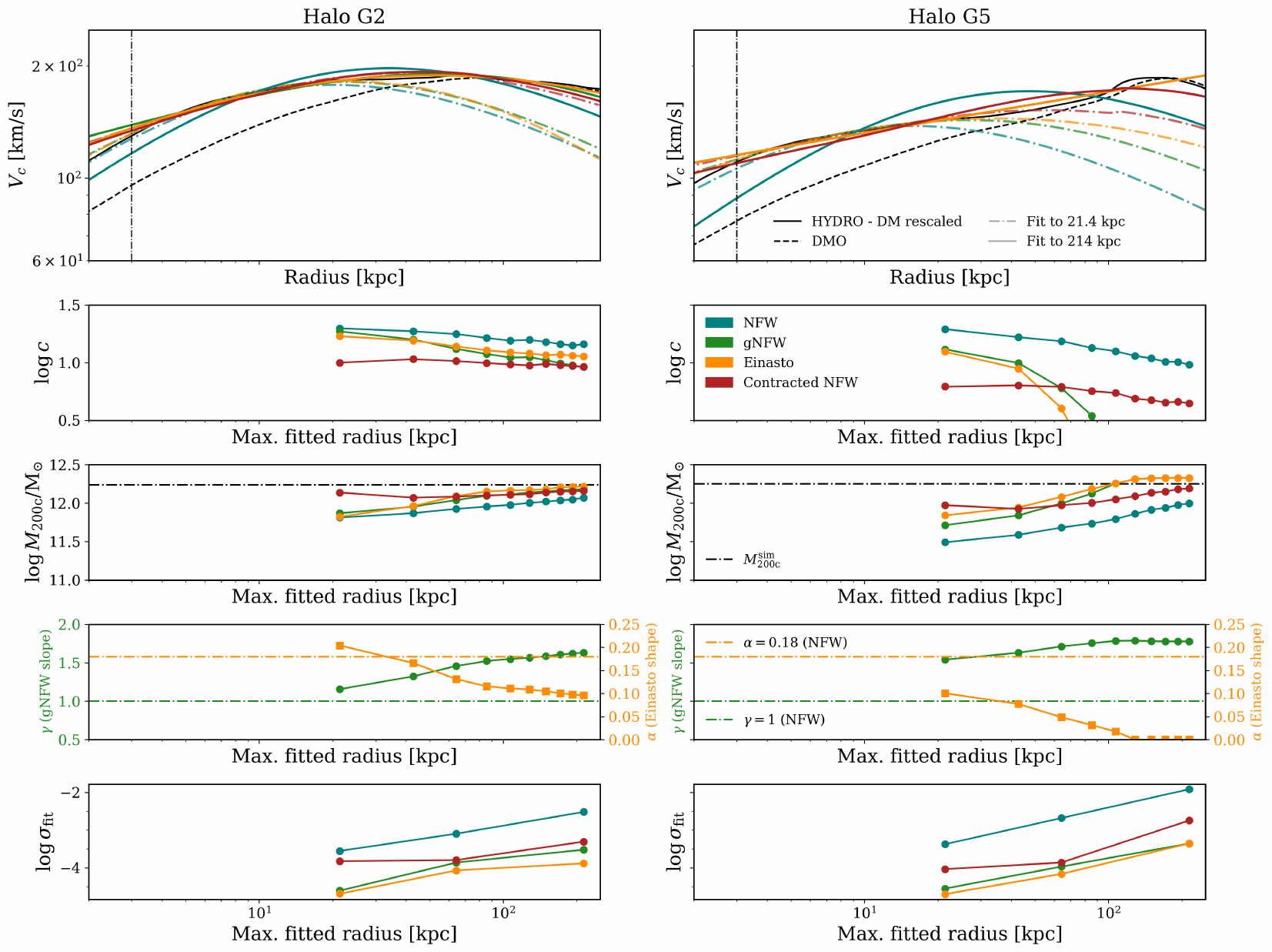}
	\caption{Comparison between a representative halo (halo G2, left) and one of the extreme outliers identified in Figure~\ref{fig:Biases on mass and concentration} (halo G5, right). \textit{First row:} Circular velocity profile of the simulated hydrodynamical halo (solid black) and its DMO counterpart (dashed black), with best-fit NFW (teal), gNFW (green), Einasto (orange), and Contracted NFW (red) profiles constrained to $r_{\rm max,fit} = 21.4$~kpc (dash-dotted) and $214$~kpc (solid). \textit{Second and third rows:} Recovered concentration and $M_{200\mathrm{c}}$ as a function of maximum fitted radius; the dash-dotted line in the third row marks the true simulated $M_{200\mathrm{c}}$. \textit{Fourth row:} Best-fit gNFW slope $\gamma$ (green, left axis) and Einasto shape parameter $\alpha$ (orange, right axis) as a function of maximum fitted radius; horizontal dash-dotted lines mark the equivalent NFW values ($\gamma=1$, $\alpha \approx 0.18$). \textit{Fifth row:} Logarithmic fit residual scatter, $\log\sigma_{\rm fit}$, as a function of maximum fitted radius. Halo G5 exhibits a localised excess in circular velocity for $r\gtrsim50$~kpc, associated with a massive subhalo near the virial radius, which drives the gNFW and Einasto shape parameters to extreme values once this radial range enters the fit, producing erratic swings in extrapolated mass and concentration despite low, well-behaved fit residuals.}
	\label{fig:appendix_outlier}
\end{figure*}

Figure~\ref{fig:appendix_outlier} compares halo G2, representative of the bulk of our sample, with halo G5, one of the extreme outliers visible in the last two columns of Figure~\ref{fig:Biases on mass and concentration}. The top row shows the simulated circular velocity profile (HYDRO - DM rescaled, solid black; DMO, dashed black) together with best-fit models (NFW, teal; gNFW, green; Einasto, orange; Contracted NFW, red) constrained to an inner radius of 21.4 kpc (dash-dotted) and to the full typical virial radius, 214 kpc (solid). For halo G2, all four models track the simulated profile closely at all radii probed in the fit. For halo G5, by contrast, the simulated profile itself departs from a smooth, monotonically-declining shape at $r \gtrsim 50~\mathrm{kpc}$, developing a localised excess in $V_{c}$ that is absent from the bulk of the sample. Cross-matching against the halo catalogue confirms that this feature coincides with the presence of a massive subhalo (or close companion) projected near the virial radius -- both in the hydrodynamical and dark matter only runs, which breaks the spherical symmetry assumed by all four fitting models.

The consequences of this localised departure are most clearly seen in the shape parameters of the flexible models (fourth row). For halo G2, the gNFW slope $\gamma$ and Einasto shape parameter $\alpha$ evolve smoothly and monotonically from their NFW-equivalent values ($\gamma = 1$, $\alpha \approx 0.18$) as the fitted range widens, consistent with the behaviour discussed in Section~\ref{sec: Fitting range sensitivity} and approaching realistic values. For halo G5, this evolution proceeds smoothly only while the fitted domain remains inside $\sim 50~\mathrm{kpc}$; once the substructure-induced feature enters the fit, $\alpha$ is driven towards zero and $\gamma$ saturates near its imposed upper bound, as both models attempt to reproduce the localised excess by adopting increasingly extreme, quasi-power-law density profiles. This directly produces the abrupt swings in concentration and mass seen in the second and third rows: $\log c$ drops sharply and $\log M_{200\mathrm{c}}$ rises sharply over the same narrow range of $r_{\rm max,fit}$ where the shape parameters saturate. The NFW and CNFW models, lacking this additional shape freedom, cannot reproduce the feature as closely, but for the same reason their extrapolated parameters remain comparatively stable and, in the case of concentration, realistic. 

Notably, the bottom row shows that this instability is not signalled by the internal fit-quality metric $\log\sigma_{\rm fit}$: for halo G5, gNFW and Einasto retain residual scatter as low as, or lower than, the well-behaved models, even in the regime where their extrapolated mass and concentration are least reliable. This reinforces the point made in Section~\ref{sec:Implications of contraction on MW} that internal diagnostics of fit quality are, on their own, poor indicators of the reliability of an extrapolation. We find this behaviour -- a localised departure from spherical symmetry driving extreme shape-parameter excursions in the most flexible models -- to be characteristic of all outliers identified in Figure~\ref{fig:Biases on mass and concentration}, and we have verified that it is insensitive to our choice of radial binning.

\section{Generalised NFW and Einasto models in the mass-concentration plane}
\label{app:gnfw_einasto_cm}

\begin{figure*}
\includegraphics[width=\textwidth]{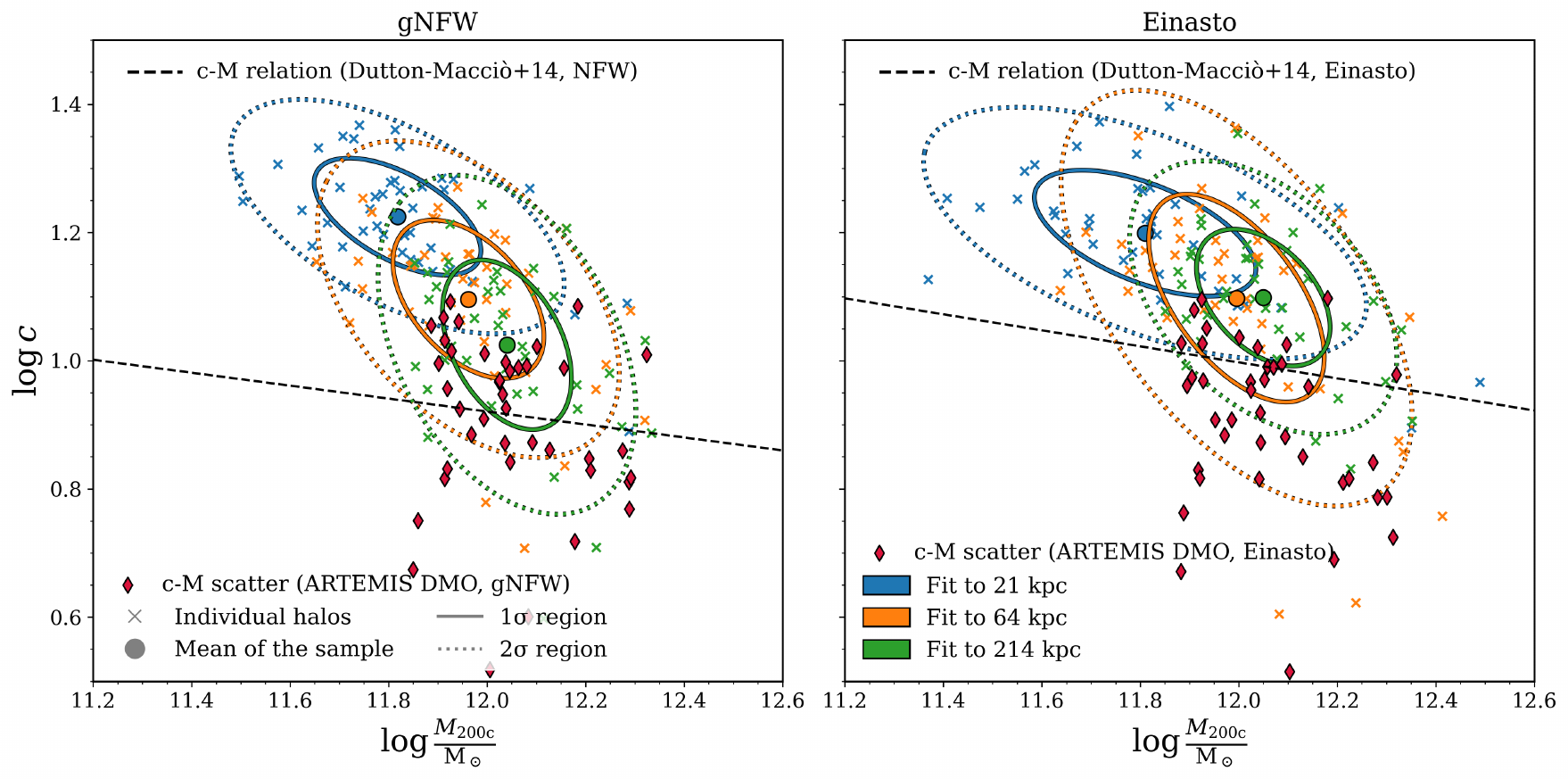}
\caption{
Recovered concentration–mass relations for gNFW (left) and Einasto (right) profile fits. Colored crosses show individual haloes for different maximum fitted radii, while large filled circles indicate the corresponding sample means. Solid and dashed ellipses denote the $1\sigma$ and $2\sigma$ covariance regions, respectively. Dark red diamonds show the intrinsic $c$--$M_{200\rm c}$ distribution of the ARTEMIS DMO haloes, obtained from fits to their full radial extent, and therefore provide the appropriate simulation benchmark. The dashed black line indicates the \citet{Dutton_2014} relation (for NFW parametrisations on the left, Einasto on the right) shown for reference.
}
\label{fig:gnfw_einasto_cm}
\end{figure*}

In this appendix, we present the behaviour of gNFW and Einasto profile fits in the $c$--$M_{200\rm c}$ plane, complementing the analysis of NFW and CNFW models discussed in Section~\ref{sec: c-M relation}. The results are shown in Fig.~\ref{fig:gnfw_einasto_cm}, constructed in the same manner as Fig.~\ref{fig:NFW-CNFW in c-M plane}.

Both gNFW and Einasto fits exhibit systematic offsets in the $c$--$M_{200\rm c}$ plane when constrained to the inner regions, consistent with the behaviour discussed in Section~\ref{sec: c-M relation}. In particular, fits restricted to $r \lesssim 20~\mathrm{kpc}$ overestimate concentrations and underestimate halo masses, displacing haloes away from the intrinsic DMO relation by amounts comparable to or exceeding the $2\sigma$ population scatter.

However, clear differences emerge between the two models. Einasto fits (right panel) lie systematically closer to both the intrinsic ARTEMIS DMO distribution and the reference $c$--$M_{200\rm c}$ relation than either gNFW or NFW fits, even when restricted to small fitting radii. In contrast, gNFW fits (left panel) display a much stronger dependence on the fitted radial range: inner-halo fits are significantly offset, while extending the fits to larger radii progressively shifts the recovered parameters towards the DMO locus.

Another notable distinction is that, unlike NFW, both gNFW and Einasto fits recover distributions that overlap with the intrinsic DMO scatter when the full radial extent of the halo is included. This indicates that the additional degree of freedom in these models can partially absorb the structural differences between hydrodynamical and DMO haloes, thereby reducing the apparent tension in the $c$--$M_{200\rm c}$ plane. However, this flexibility does not eliminate the underlying bias when only inner-halo data are available; instead, it redistributes the mismatch across parameters. This effect is particularly pronounced for gNFW fits, where strong degeneracies between shape parameters lead to increased scatter and sensitivity to the fitted radial range.

Overall, these results highlight that increased model flexibility can mitigate discrepancies with the DMO $c$--$M_{200\rm c}$ relation when wide radial coverage is available, but does not provide a physically robust correction for baryon-induced contraction. In contrast, the CNFW model achieves consistency with the expected scaling by explicitly accounting for these effects, rather than effectively marginalising over them through additional degrees of freedom.

\bsp
\label{lastpage}
\end{document}